\documentclass[journal=jacsat,manuscript=article]{achemso}

\usepackage{chemformula} 
\usepackage{setspace} 
\usepackage[T1]{fontenc} 
\usepackage{csvsimple}
\usepackage{graphicx}
\usepackage{subcaption}
\usepackage{subfig}
\usepackage{amsmath}
\usepackage{amsfonts}
\usepackage{booktabs}
\usepackage{placeins}
\usepackage{nameref}
\usepackage{array}
\usepackage{caption}
\usepackage[ruled,vlined]{algorithm2e}

\author{Parastoo Semnani}
\affiliation[TU Berlin]{Machine Learning Group, TU Berlin, Berlin, Germany}
\alsoaffiliation[BIFOLD]{Berlin Institute for the Foundations of Learning and Data, Berlin, Germany}
\alsoaffiliation[BASLEARN]{BASLEARN–TU Berlin/BASF Joint Lab for Machine Learning, TU Berlin, Berlin, Germany}

\email{p.semnani@tu-berlin.de}
\author{Mihail Bogojeski}
\affiliation[TU Berlin]{Machine Learning Group, TU Berlin, Berlin, Germany}
\alsoaffiliation[BIFOLD]{Berlin Institute for the Foundations of Learning and Data, Berlin, Germany}

\author{Florian Bley}
\affiliation[TU Berlin]{Machine Learning Group, TU Berlin, Berlin, Germany}
\alsoaffiliation[BIFOLD]{Berlin Institute for the Foundations of Learning and Data, Berlin, Germany}

\author{Zizheng Zhang}
\author{Qiong Wu}
\author{Thomas Kneib}
\affiliation[Georg-August-University Göttingen]{Chair of Statistics and Campus Institute Data Science, Georg-August-University Göttingen, Göttingen, Germany}

\author{Jan Herrmann}
\author{Christoph Weisser}
\author{Florina Patcas}
\affiliation[Production AI]{BASF SE, Ludwigshafen, Germany}

\author{Klaus-Robert Müller}
\affiliation[TU Berlin]{Machine Learning Group, TU Berlin, Berlin, Germany}
\alsoaffiliation[BIFOLD]{Berlin Institute for the Foundations of Learning and Data, Berlin, Germany}
\alsoaffiliation[Max Planck]{Max Planck Institute for Informatics, Saarbrücken, Germany}
\alsoaffiliation[Korea University]{Department of Artificial Intelligence, Korea University, Seoul, South Korea}
\email{klaus-robert.mueller@tu-berlin.de}

\title[An \textsf{achemso} demo]
  {A Machine Learning and Explainable AI Framework Tailored for Unbalanced Experimental Catalyst Discovery}


\begin{document}







\begin{abstract}

The successful application of machine learning in catalyst design depends on high-quality and diverse data to ensure effective generalization to novel compositions, thereby aiding in catalyst discovery.
However, due to the complex interactions of catalyst components, the design of novel catalysts has long relied on trial-and-error, a costly and labor-intensive process that results in scarce data that is heavily biased towards undesired, low-yield catalysts. Despite the increasing popularity of machine learning applications in this field, most of the efforts so far have not focused on dealing with the challenges presented by such experimental data. 
To address these challenges, we introduce a robust machine learning and explainable AI framework to accurately classify the catalytic yield of various compositions and identify the contributions of individual components to the yield. This framework combines a series of ML practices designed to handle the scarcity and imbalance of catalyst data.
We apply the framework to the task of determining the yield of different catalyst combinations in oxidative methane coupling, and use it to evaluate the performance of a range of ML models: tree-based models (such as decision trees, random forest, and gradient boosted trees), logistic regression, support vector machines, and neural networks.
These experiments demonstrate that the methods used in our framework lead to a significant improvement in the performance of all but one of the evaluated models.
Additionally, the decision-making process of each ML model is analyzed by identifying the most important features for predicting catalyst performance using explainable AI (XAI) methods. Our analysis found that XAI methods, which provide class-aware explanations, such as Layer-wise Relevance Propagation, managed to identify key components that contribute specifically to high-yield catalysts. These findings align with chemical intuition and existing literature, reinforcing their validity. We believe that such insights can assist chemists in the development and identification of novel catalysts with superior performance. Illustration of the abstract is depicted in Figure~\ref{fig:abstract_fig.png}.

\end{abstract}

\begin{figure}[ht]
    \centering
    \includegraphics[width=0.9\textwidth]{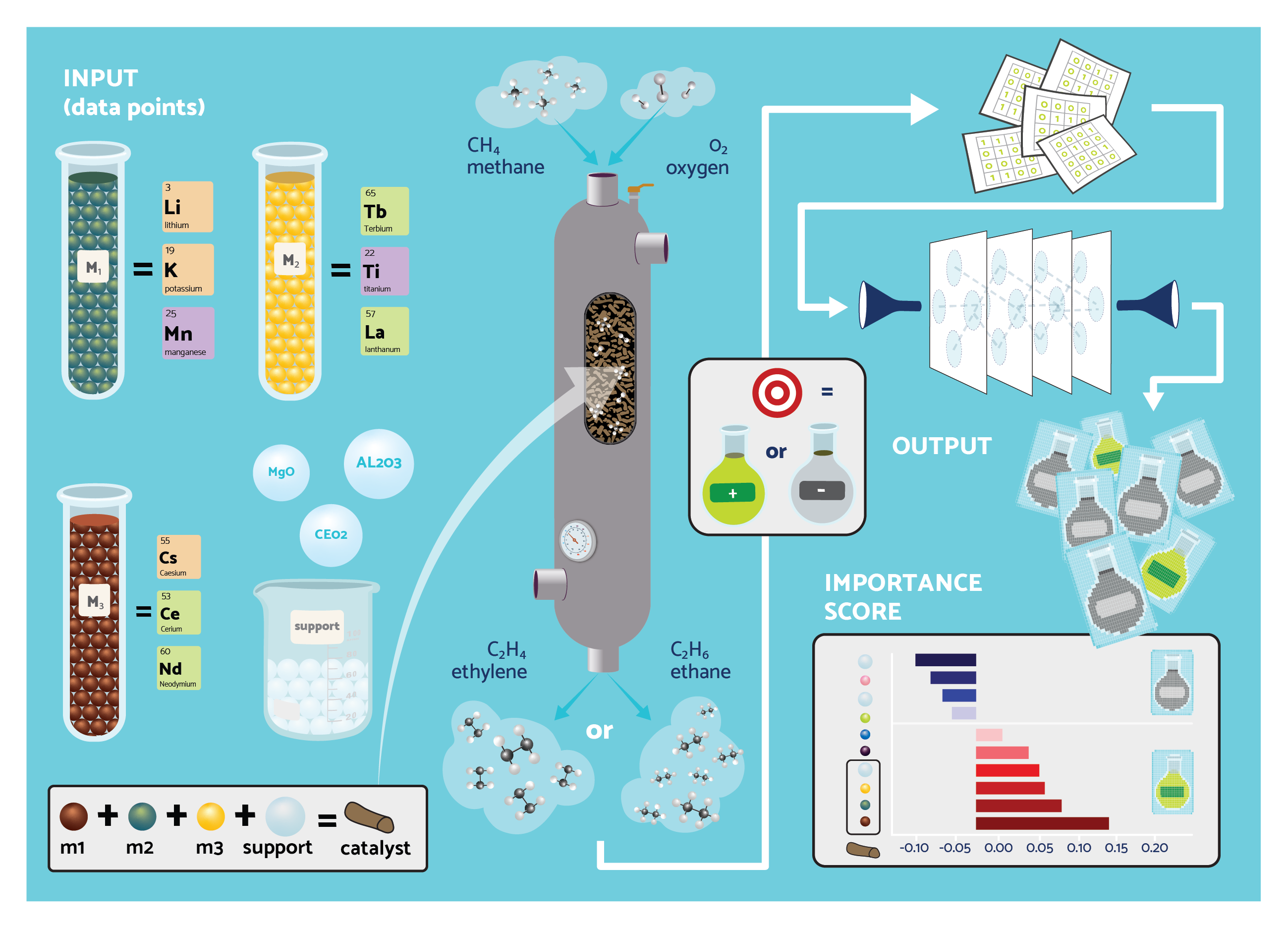}
    \caption{Visual abstract of ML-guided catalyst design: The figure illustrates the process of oxidative methane coupling, where the catalyst consists of M1-M2-M3/support material. This catalyst is tested in high-throughput screening to determine the yield of each composition. The resulting data is then utilized in various machine learning models, whose performance and feature importance are subsequently analyzed.}
    \label{fig:abstract_fig.png}
\end{figure}


\section{Introduction}
Machine learning (ML) models have recently become popular in the field of heterogeneous catalyst design~\cite{graser2018machine, andersen2019beyond, ma2020machine, bogojeski2021forecasting, margraf2023exploring}. The inherent complexity of the interactions between catalyst components is very high, leading to both synergistic and antagonistic effects on catalyst yield that are difficult to disentangle. Therefore, the discovery of well-performing catalysts has long relied on serendipitous trial and error~\cite{vojvodic2015new, goldsmith2018machine}. 
 
Unlike traditional methods based on simplified models and heuristics, ML methods excel at identifying complex patterns and non-linear relationships between various catalyst components. This capability is particularly advantageous in catalyst design, where ML can offer insights into nuanced component interactions, crucial for optimizing yield~\cite{rangarajan2024artificial, suzuki2019statistical, wang2023interpretable}.
However, the application of ML methods in catalyst design faces several challenges. Most prominent challenge is the scarcity of large and unbiased datasets. Despite significant efforts in data acquisition and curation~\cite{nguyen2021learning, nguyen2019high, zavyalova2011statistical, takahashi2022synthesis, nishimura2022high}, datasets often remain small and limited due to the high costs in labor and time.
Moreover, existing data often favors historically successful or easily testable catalysts, leading to bias in the choice of elements and supports~\cite{toyao2019machine, jia2019anthropogenic, brown2015understanding}. 

These limitations restrict a model's exposure to a wide range of catalytic scenarios, leading to biases that can prevent the accurate learning of underlying relationships between catalyst components, thus reducing model's generalization performance~\cite{lapuschkin2019unmasking}. Therefore, additional research is needed to complement existing approaches and improve the robustness of ML models, enabling effective learning and generalization with scarce and biased data.

To tackle these challenges, we propose a robust ML and explainable AI (XAI) framework designed to handle the scarcity and imbalance of catalyst data (see Figure~\ref{fig:ML framework.png}). This study contributes to ML-guided catalyst design by implementing our framework on the unbiased dataset provided by~\citeauthor{nguyen2021learning}, on the oxidative methane coupling (OCM), which includes a diverse selection of elements and supports. Our approach accurately classifies catalytic yields and identifies the contributions of individual components to the yield, using performance metrics and sampling strategies tailored to small and imbalanced datasets.
These metrics address class imbalance to enhance performance estimation reliability, while the sampling methods mitigate biases and the impacts of over-represented classes during training. Integrating these approaches within an evaluation and explanation framework strengthens model robustness for reliable predictions. Furthermore, we systematically assess a variety of ML models within this framework, highlighting their capabilities and limitations. Recognizing the necessity for model interpretability in catalysis, we apply XAI methods~\cite{bach2015pixel, montavon2018methods, arrieta2020explainable, samek2021explaining, minh2022explainable, saranya2023systematic} to analyze strongly non-linear models, such as neural networks and support vector machines (SVM), identifying key features that influence their decisions and providing insights into their decision-making processes.

In summary, this work proposes more robust performance metrics and sampling strategies, explores a diverse set of ML models, and applies XAI methods to analyze their decisions. We aim to pave the way for effective ML-guided catalyst design under data scarcity, explain the importance of catalyst components, and thus accelerate the efficient design of experiments and material discovery.

\section{Materials and methods}\label{sec:mat_and_met}

\subsection{Data}\label{sec:data}

Under the effects of certain catalysts, OCM converts methane to \ch{C2} products, e.g. \ch{C2H4} and \ch{C2H6}, which serve as the fundamental building blocks in the chemical industry. Thus, the effectiveness of a catalyst is often measured by the percentage of \ch{C2} yield. Researchers have applied catalyst informatics to OCM, using data analysis and ML methods to identify synergistic combinations like Na-La, Na-Mn, and Ba-Sr~\cite{zavyalova2011statistical}. Current challenges include inconsistent experimental methods and biases in component choices among different publications \cite{schmack2019meta,jia2019anthropogenic}.

To address these challenges,~\citeauthor{nguyen2021learning} have gathered unbiased and process-consistent OCM data via a high-throughput screening (HTS) instrument for 300 quaternary structured catalysts, with each component being randomly selected from a predefined range of candidates. The quaternary structure of the catalyst, M1-M2-M3/support, consists of three active elements (M1-M2-M3) randomly selected from 28 commonly used elements (including 'none' as an option) with replacement, and one support randomly selected from 9 oxides. To ensure unbiased selection, 300 combinations are randomly chosen as candidate catalysts from all possible combinations. Evaluation experiments for each candidate catalyst under 135 different reaction conditions are then performed via HTS. Specifically, one combination of temperature, input ratios, total flow, and atmospheric pressure defines one reaction condition. 
Only the data of one reaction condition with the highest \ch{C2} yield is recorded for each candidate catalyst. Apart from \ch{C2} yield, another two quantities \ch{CH4} conversion and \ch{C2} selectivity are recorded. \ch{CH4} conversion measures how much of the input methane is converted. \ch{C2} selectivity measures how much of the output is the desired output, i.e. \ch{C2} products. Thus, the product of \ch{CH4} conversion and \ch{C2} selectivity equals \ch{C2} yield, which also indicates the conversion-selectivity trade-off. There are, in total, 291 records for individual catalysts in the dataset, since the performance scores of 9 catalysts are missing.~\citeauthor{nguyen2021learning} provides informative interpretations from a chemical perspective based on the statistical analysis of the experimental data. 

To facilitate the efficient discovery of combinatorial catalysts,~\citeauthor{nguyen2021learning} have prepared the data with an unbiased selection of elements, making it potentially beneficial for ML applications. The target variable is set as the best \ch{C2} yield, which is a binary variable that is set as true if the yield is larger than 13\% and false when the yield is lower. The dataset consists of 51 high-yield catalysts and 240 low-yield catalysts in total. The data consists of 49 boolean features denoting the presence of elements (27), supports (9), and periodic table groups (13) in a given catalyst combination. Due to its diverse and bias-free construction, we chose to use this dataset as an example for our proposed framework for training, evaluation, and explainable AI.

While we strongly appreciate the efforts of~\citet{nguyen2021learning} in curating this unbiased dataset and making it publicly available, it is essential to highlight certain characteristics and potential issues of this data to provide context for our analysis.

First, it is important to note that the dataset only includes the optimal operation conditions for the catalyst material. As a consequence, the test conditions, such as temperature and Gas Hourly Space Velocity, serve as identifiers for each data instance rather than features that can be used for training and analysis. Additionally,~\citeauthor{nguyen2021learning} highlights the process sensitivity of the OCM reaction, indicating that test conditions may have a more profound impact on catalyst performance than changes in material composition.

Finally, we have found that the features denoting whether an element belongs to a specific group in the periodic system are superfluous, as they do not seem to improve the overall performance of the models when accounting for class imbalance.
Additionally, they make it more difficult to disentangle feature importance attribution from explainability methods since they correlate strongly with the elements belonging to the group. This is especially the case with groups 3, 5, 7, 8, 9, 11, and 12, which have only 1 element each, respectively \ch{Y}, \ch{V}, \ch{Mn}, \ch{Fe}, \ch{Co}, \ch{Cu}, and \ch{Zn}, resulting in the corresponding features being fully correlated. A more detailed discussion on this can be found in Supplementary Section~\nameref{sup:group_feats}.

In supervised ML, the overarching goal is to develop models that generalize well to unseen data. However, the exclusive use of optimal test conditions within the dataset poses a challenge to the model's generalization ability. By training solely on data characterized by optimal process conditions, the models may struggle to accurately predict the target variable for unseen cases subject to different sets of process conditions.

\subsection{Performance evaluation metrics}
\begin{figure}[ht]
    \centering
    \includegraphics[width=0.9\textwidth]{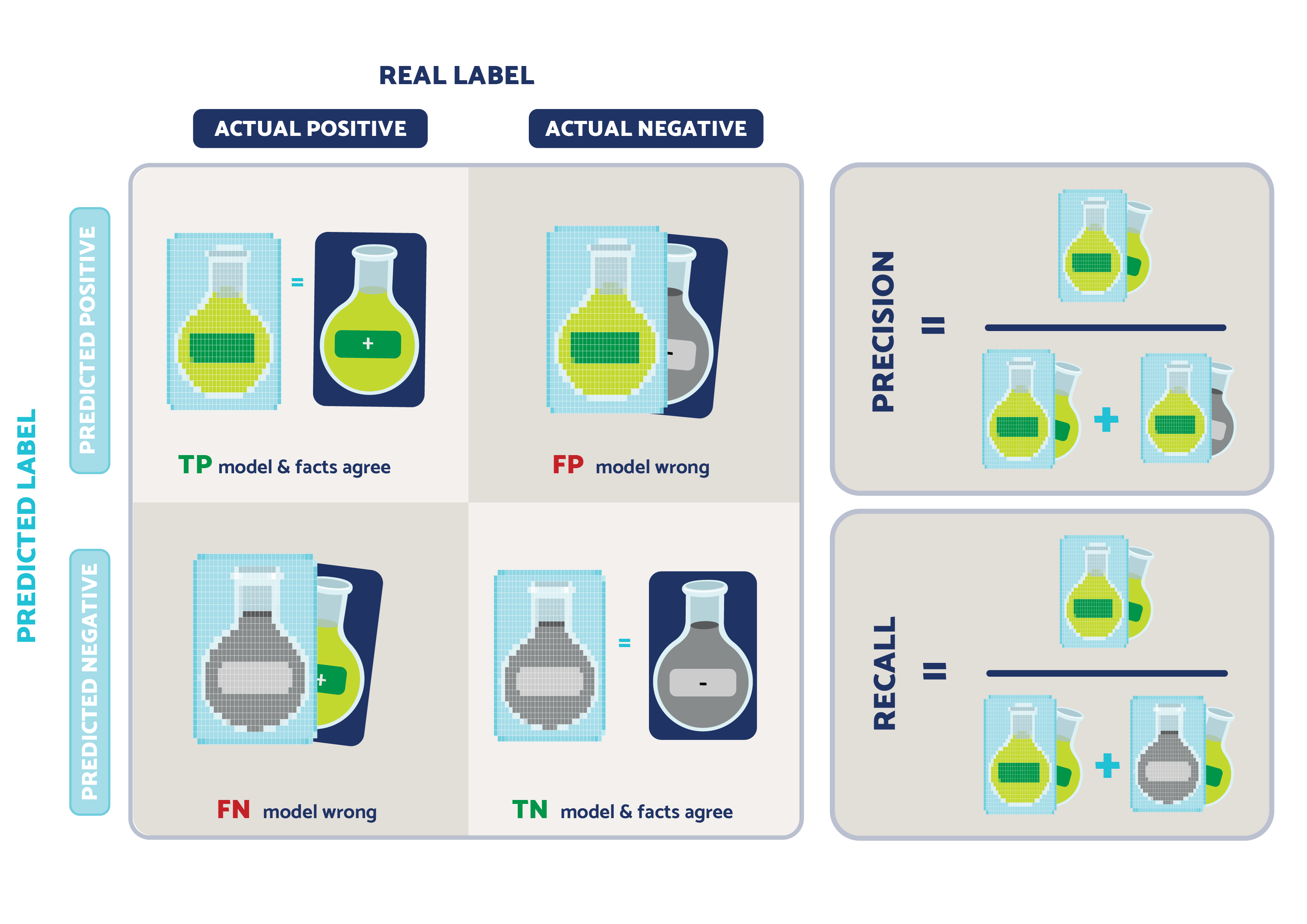}
    \caption{Illustration of evaluation metrics - The blurred symbol is the model's prediction, and the unblurred symbol is the true label of the data.}
    \label{fig:Evaluation metrics.png}
\end{figure}

Among the various measures used to evaluate the performance of ML models, this study focuses on accuracy, precision, recall, and F1-score. 
These metrics measure different aspects of model performance, each suited to different objectives and contexts. The equations for all the measures we use in this work are shown in Table~\ref{tab:metrics}, and an illustration comparing the relationships between the performance measures can be found in Figure~\ref{fig:Evaluation metrics.png}.

\emph{Accuracy} is one of the most widely used evaluation metrics for ML models, which simply measures the proportion of correct predictions, encompassing both true positives and true negatives in a single metric. It is particularly suitable for balanced datasets, where the number of samples in each class is roughly equal. In the case of highly unbalanced class ratios, accuracy can be misleading since a classifier only predicting the majority class would still be able to achieve high accuracy.

To overcome this shortcoming, other evaluation measures have been introduced that better reflect the different aspects of the problem. In the case of catalyst design, we are more interested in one of the two classes, namely high-yield catalysts. This is why the measures precision, recall and F1-score are especially relevant here. \emph{Precision} measures the proportion of true positive predictions among all positive predictions and is valuable when the cost of false positives is high.
\emph{Recall}, also known as sensitivity or true positive rate, gauges the ability of the model to capture all positive samples and is crucial when the cost of false negatives is a concern. It is defined as the ratio of true positives to the sum of true positives and false negatives. Finally, the \emph{F1-score} is the harmonic mean of precision and recall and is particularly useful when the class ratios are imbalanced, and the positive class is especially important, which is exactly the case in our catalyst yield classification task.

\begin{table}[htpb]
    \centering
    \renewcommand{\arraystretch}{3} 
    \begin{tabular}{|>{\centering\arraybackslash}p{5cm}|>{\centering\arraybackslash}p{5cm}|}
        \hline
        $\text{Accuracy} = \frac{\text{TP} + \text{TN}}{\text{TP} + \text{FP} + \text{TN} + \text{FN}}$ & $\text{Precision} = \frac{\text{TP}}{\text{TP} + \text{FP}}$ \\
        \hline
         $\text{Recall} = \frac{\text{TP}}{\text{TP} + \text{FN}}$ & $F1 = 2 \cdot \frac{\text{Precision} \cdot \text{Recall}}{\text{Precision} + \text{Recall}}$ \\
        \hline
    \end{tabular}
    \caption{Definition of different commonly used performance evaluation metrics for ML models. TP denotes the number of true positives, TN that of true negatives, FP denotes the number of false positives, and FN the number of false negatives.)}
    \label{tab:metrics}
\end{table}
\vspace{-1em}

\subsection{Resampling}

ML models often struggle within scenarios with highly imbalanced class distributions~\cite{he2013imbalanced, fernandez2018learning}. Because most ML models are designed for datasets with an equal number of observations for each class, if the imbalance is not accounted for, the models may prioritize the majority class and overlook the minority class, negatively impacting overall performance.

One common approach to addressing this issue is by employing resampling techniques, which use various strategies to oversample the minority class or undersample the majority class in order to balance the dataset~\cite{tomek1976two, kubat1997addressing, ling1998data,drummond2003c4,liu2007generative,huang2016learning,bellinger2020remix,muttenthaler2023set}.

We choose to perform oversampling using Synthetic Minority Oversampling Technique (SMOTE)~\cite{chawla2002smote}, and following the recommendation of~\citet{chawla2002smote}, we combine SMOTE with random undersampling of the majority class.

\subsection{Cross-validation}
When dealing with small datasets, the performance of the model can depend quite strongly on the choice of the training and test subsets, making it difficult to obtain a reliable estimate of the model's generalization error. In such cases, providing an accurate and unbiased estimate of the error through cross-validation (CV) and hyper-parameter tuning becomes essential, which in turn allows for the selection of the most robust and best-performing model~\cite{hastie2009elements, toyao2019machine}.

In our study, we use a variant of nested $k$-fold cross-validation to reliably evaluate model performance on unseen data~\cite{muller2001introduction, hansen2013assessment}. In $k$-fold cross-validation, the dataset is divided into k-subsets of roughly equal size, one of which is chosen as the validation set, another one as the test set, and the rest are combined into the training set. The model is then trained using the training set, while the validation set is used to select the best-performing set of hyperparameters during training. Finally, the test set is used to evaluate the model's predictive power on unseen data. This procedure is then performed for different allocations of the subsets to the training, validation, and test datasets. Nested $k$-fold cross-validation improves robustness by creating multiple different random splits of the dataset into k subsets, and performing the whole process of $k$-fold cross-validation multiple times. Using this procedure, we ensure that each data point is represented in the train, validation, and test set in different splits, preventing overfitting and ensuring unbiased performance evaluation that is not dependent on the initial partitioning of the data.

\subsection{Machine learning models}

To showcase the general nature of our framework and provide a broad overview of the diverse approaches in machine learning, we evaluate a variety of ML models commonly used in classification tasks.

This includes a series of models from the family of tree-based models such as \emph{decision trees} (with both prepruning and postpruning), as well as \emph{random forests} and \emph{gradient boosted trees}, which are ensembles constructed of many individual decision trees. We also include logistic regression, one of the oldest and most popular methods for binary classification.

Finally, we evaluate support vector machines (SVMs) and neural networks, two powerful and highly non-linear ML methods. Detailed explanations of these models can be found in Supplementary Section~\emph{\nameref{sec:ml_models_theory}}.

\subsection{Explainable AI}
Explainable AI (XAI) techniques are playing an increasingly important role in various domains, including catalyst research.~\cite{molnar2020interpretable} While there are many approaches to explaining the ML model decisions, in this paper, we focus on XAI methods that assign importance to each input feature based on how relevant they were to the models' prediction.~\cite{arrieta2020explainable} In the catalyst design scenario explored here, such XAI methods would point out which components contribute particularly strongly to a catalyst being classified as either high- or low-yield. When the relevant components highlighted by the explanation method align with chemists' intuition, it serves as validation that the models are focusing on chemically relevant features rather than artifacts, thereby increasing confidence in their predictions.
In some cases, XAI can also yield insights into the previously unknown relations between different catalyst components and thus aid in the efficient discovery of new catalysts.

\subsubsection{Feature Importance for tree-based models}

For the three variants of decision trees, feature importance was determined based on the mean decrease in impurity across all decision nodes. This metric quantifies the contribution of each input feature to reducing impurity when splitting the data along this feature during the training process~\cite{breiman2001random}, commonly measured by the Gini index or entropy.
\begin{equation}
\label{eq:gini}
\mathrm{Gini}(t) = 1 - \sum_{i=1}^{J} p_{i}^{2},
\end{equation}
where $J$ is the total number of classes in the dataset, and $p_i$ represents the proportion of samples belonging to class $i$ at node $t$.

In this case, since the decision tree models were trained using the Gini index, we also use this as the measure of feature importance. Higher importance scores indicate a greater impact on impurity reduction, highlighting the significance of these features in the classification process.
For the random forest models, feature importance is determined by aggregating the reduction in Gini impurity achieved by splitting each feature across all trees within the ensemble:
\begin{equation}
FI_{RF}(\mathbf{x}_d) = \frac{1}{T}\sum_{t}^{T}\sum_{s}^{S_t} \Delta \mathrm{Impurity}_{t,s}(\mathbf{x}_d).
\end{equation}
Here, $\mathbf{x}_d$ is the $d$-th feature of the input vector $\mathbf{x}$, $T$ is the total number of trees in the random forest, $S_t$ is the number of splits in tree $t$, and $\Delta \mathrm{Impurity}_{t,s}(\mathbf{x}_d)$ is the decrease in Gini impurity attributable after split $s$ in tree $t$, if feature $\mathbf{x}_d$ was used~\cite{breiman2017classification}.

In XGBoost, a feature's importance increases with its contribution to splits during tree construction and is calculated by summing the gain (see Supplementary Section~\emph{\nameref{sec:xgb_theory}}) of each specific feature across all trees and splits:

\begin{equation}
FI_{XGB}(\mathbf{x}_d) = \sum_{s}^{S} \mathrm{Gain}_s(\mathbf{x}_d),
\end{equation}
where $\mathbf{x}_d$ once again refers to the feature $d$ in the input $\mathbf{x}$, S is the total number of splits across all trees, and $\mathrm{Gain}_s(\mathbf{x}_d)$ is the gain resulting after split $s$, if feature $\mathbf{x}_d$ was used for this split.

\subsubsection{Layer-wise Relevance Propagation (LRP)}\label{sec:lrp}

LRP is a popular explaining technique for interpreting predictions of complex neural network models in terms of latent and input features~\cite{bach2015pixel, montavon2018methods, samek2021explaining}. In contrast to feature importances for tree-based models, which primarily explain the parameters of the model itself, LRP produces local explanations for the classification of each sample. Using so-called propagation rules \cite{Montavon2019_overview}, LRP assigns a relevance value to each neuron by iteratively backpropagating the model output through the network layers until the input layer is reached. Propagation rules are chosen to be conservative, meaning that total relevance in each layer is equivalent to the network output. In general, most LRP rules compute lower-layer relevance $R_i$ given upper-layer relevance $R_j$ using the following generic format:
\begin{equation}\label{eq:lrp_general}
    R_i = \sum_{j} \frac{\rho(w_{ij}) \cdot a_i}{\sum_{0, i'} \rho(w_{i'j}) \cdot a_{i'} + \epsilon} \cdot R_j.
\end{equation}
In the above formulation, the sum $ \sum_{j}$ runs over upper-layer neurons $\{a_j\}_j$, whereas the sum $\sum_{0, i'}$ runs over lower-layer neurons $\{a_i'\}_i'$ including the bias represented as the additional neuron $a_0$. The variable $w_{ij}$ describes the weight connecting the lower-layer neuron activation $a_i$ and the upper-layer neuron $a_j$, while $\rho$ describes some functional dependence of the neuron weights.  To avoid division by zero, most LRP rules stabilize the above denominator adding a small positive value $\epsilon$. As exemplified by the above formula, most propagation rules distribute relevance depending on how much each lower-layer neuron has contributed to the output of the higher-layer neuron. Contrary to feature importance explanations, LRP relevance values can be either positive or negative, thus describing how much a given feature attributed to the model deciding in favor of one class or the other.

To start relevance propagation, a suitable neuron output must be chosen to set upper-layer relevance. One possible set of explained neurons is the neurons in front of the final softmax layer, which aggregate evidence for a given class. The upper-layer evidence neurons form a linear layer and compute activations for a given class $c$ as follows:
\begin{equation}
    a_c = \sum_{0,  k} w_{c, k} \cdot a_k.
\end{equation}
However, as it has been found that explaining only one class-evidence neuron does not contextualize evidence of competing classes, an alternate approach is to explain the logit of class probabilities instead \cite{Montavon2019_overview}. This quantity is expressed as follows:
\begin{equation}
    \eta_c = \frac{\log(p_c)}{\log(1-p_c)}.
\end{equation}
In a two-class setting with class indices $1$ and $-1$, this further simplifies as follows:
\begin{equation}
    \eta_1 = a_1 - a_{-1}
\end{equation}
Combining the evidence weight vectors, $\eta$ can then be expressed as the following explainable neuron:
\begin{equation}
    \eta_1 = \sum_{ 0, k} (w_{1, k} - w_{-1, k}) \cdot a_k.
\end{equation}
This neuron $\eta$ can then finally be used as the starting point of the relevance propagation procedure for the classifier. To explain subsequent Multi-Layer Perceptron (MLP) layers, we applied the $\gamma$-rule, which sets the functional dependence $\rho(w_{ij}) = w_{ij} + \gamma \cdot \max(0, w_{ij})$ given a value of $\gamma$, that we set to 0.2. The $\gamma$-rule emphasizes positive contributions to neuron outputs, which has been shown to improve the stability and faithfulness of the resulting explanation~\cite{Montavon2019_overview}.

While the LRP rules conserve relevance, some relevance in each layer gets assigned to neuron bias terms, which cannot be explained in terms of input features. Consequently, the relevance assigned to the input does not perfectly match the class evidence of the prediction. To account for the relevance lost to the biases and improve the interpretability of the relevances assigned to the input, we rescale the input feature relevance values $\{R_d\}_d$ such that positive relevance adds up to the positive-class evidence and vice versa.

The rescaling is done using sign-dependent factors $\rho^{+}$ and $\rho^{-}$, which are determined based on the positive and negative contributions to the output neuron $\eta_1$. More specifically, for each sample, the positive input relevances are scaled such that their sum is equal to the positive contributions to $\eta_1$, and vice-versa for the negative relevance: 

\begin{align}\label{eq:nn_rel_rescaling}
    \sum_d \rho^{+} \cdot \max(R_d, 0) =& \; \sum_{0, k} \max\left((w_{1, k} - w_{-1, k}) \cdot a_k, 0\right), \\
    \sum_d \rho^{-} \cdot \min(R_d, 0) =& \; \sum_{0, k} \min\left((w_{1, k} - w_{-1, k}) \cdot a_k, 0\right).
\end{align}

This rescaling strategy ensures that the relevance in the inputs conserves the value of the output neuron $\eta_1$ in a way that preserves the original sign of the input relevances.

\subsubsection{LRP for neuralized SVMs}\label{sec:lrp_for_neuralised_svms}

By default, LRP requires a neural network structure and is, without further modification, not suited to explain kernel-based models. To overcome this limitation and provide faithful explanations with LRP, \citet{kauffmann2022clustering} introduced the concept of neuralization. Neuralization transforms a kernel-based model into a neural network structure producing equivalent decisions explainable with propagation-based XAI methods.

In the case of RBF-SVMs,~\citet{XAI_SVM_thesis} modified the SVM predictive function $f(\textbf{x})$ of Eq. \ref{eq:svm_pred_fun} in the following way:

\begin{equation}\label{eq:g_of_x_ansatz}
    g(\mathbf{x}) = \log \left(\sum_{i} \alpha_i \exp(-\gamma \cdot\lVert \mathbf{x - x}_i \lVert^2)\right) - \log \left(\sum_j \lvert\alpha_j\rvert \exp(-\gamma\cdot \lVert \mathbf{x - x}_j \lVert^2)\right)
\end{equation}
Here, the index $i$ runs over positive-class support vectors and $j$ over the negative-class support vectors. Therefore, $\mathbf{x}_i$ and $\mathbf{x}_j$ describe the support vectors themselves, and $\alpha_i$ and $\alpha_j$ are the associated dual coefficients of the SVM. The two logarithmic terms can be interpreted as evidence for the two competing classes.
The transformed classifier $g(\textbf{x})$ is guaranteed to produce an equivalent classification to the original SVM. The authors went on and transformed $g(\textbf{x})$ into the following neural network structure:
\begin{equation}\label{equation_svm_neuralisation}
\begin{aligned}
    g(\mathbf{x}) &= \gamma \cdot \underset{j}{\text{min}}^{\gamma} \left(\underset{i}{ \text{max}}^\gamma \left( w_{ij}^T \cdot \mathbf{x} + b_{ij}\right)\right)\\
    \text{where} \quad w_{ij} &= 2 \cdot (\mathbf{x}_i - \mathbf{x}_j), \\
     \quad b_{ij} &= \lVert \mathbf{x}_j \rVert^2 - \lVert \mathbf{x}_i \rVert^2 + \frac{1}{\gamma} \cdot \log\left(\frac{\alpha_i}{\lvert\alpha_j\rvert}\right) \\
\end{aligned}
\end{equation}
The above formulation utilizes the soft-min and soft-max pooling-layer definitions of ~\citet{kauffmann2022clustering}, where $\min^\gamma(\cdot)$ is defined as $-\frac{1}{\gamma} \log \sum \exp(-\gamma \cdot (\cdot))$, and $\max^\gamma(\cdot)$ is defined as $\frac{1}{\gamma} \log \sum \exp(\gamma \cdot (\cdot))$. Thus, the neuralized RBF-SVM can be summarized as two pooling layers preceded by one detection layer with one detection neuron for each pair $ij$ of positive-class and negative-class support vectors.

To propagate relevance through the first two pooling layers, we follow the approach of ~\citet{kauffmann2022clustering}. Based on the concept of Deep Taylor Decomposition \cite{Montavon2019_overview}, the authors derived the following conservative propagation rules for the soft-min and soft-max layers:
\begin{align}
        R_j &= \frac{\exp(-a_j)}{\sum_{j'} \exp(-a_{j'})} \cdot R_k \label{eq:min_takes_most}\\
        R_{ij} &= \frac{\exp(a_{ij})}{\sum_{i'} \exp(a_{i'j})} \cdot R_j\label{eq:max_takes_most}
\end{align}
To propagate through the linear layer and produce input feature relevance values $\{R_d\}_d$, we use the LRP-$0$ rule, which attributes relevance according to the element-wise product of the neuron weights and input features:

\begin{equation}
    R_d = \frac{w_{ij, d} \cdot \mathbf{x}_d}{\sum_{0, d'} w_{ij, d'} \cdot \mathbf{x}_{d'}} \cdot R_{ij}
\end{equation}
During this propagation, however, relevance is naturally lost in the linear layer to the biases. To compensate for lost relevance and ensure interpretability, we reweight relevance such that positive and negative relevance add up to the original class evidence of the explained model in Eq. \ref{eq:g_of_x_ansatz}. In particular, we identify sign-dependent reweighting factors $\rho^{+}$ and $\rho^{-}$ to rescale feature relevance such that the following relations hold:
\begin{align}\label{eq:svm_rel_rescaling}
    \sum_d \rho^{+} \cdot \max(R_d, 0) =& \; \log \left( \sum_i \alpha_i \exp(-\gamma \cdot\lVert \mathbf{x - x_i} \rVert^2)\right), \\
    \sum_d \rho^{-} \cdot \min(R_d, 0) =& \;-\log \left( \sum_j \alpha_j \exp (-\gamma \cdot \lVert \mathbf{x - x_j} \rVert^2 )\right).
\end{align}

\section{Results and discussion}
To ensure the reliability and best practices for evaluating ML methods in catalyst design, we elaborate on our proposed ML framework tailored for datasets characterized by small-scale and class imbalances. This section provides an overview of the framework's architecture and its application to various ML models. We then assess the impact of different framework components (e.g., performance measures, resampling techniques) on model performance. Finally, we leverage XAI techniques to analyze the most relevant features identified by each model and investigate common features across models to gain an understanding of the underlying data.

\subsection{Evaluation framework}\label{sec:evaluation_framework}
\begin{figure}[ht]
    \centering
    \includegraphics[width=0.9\textwidth]{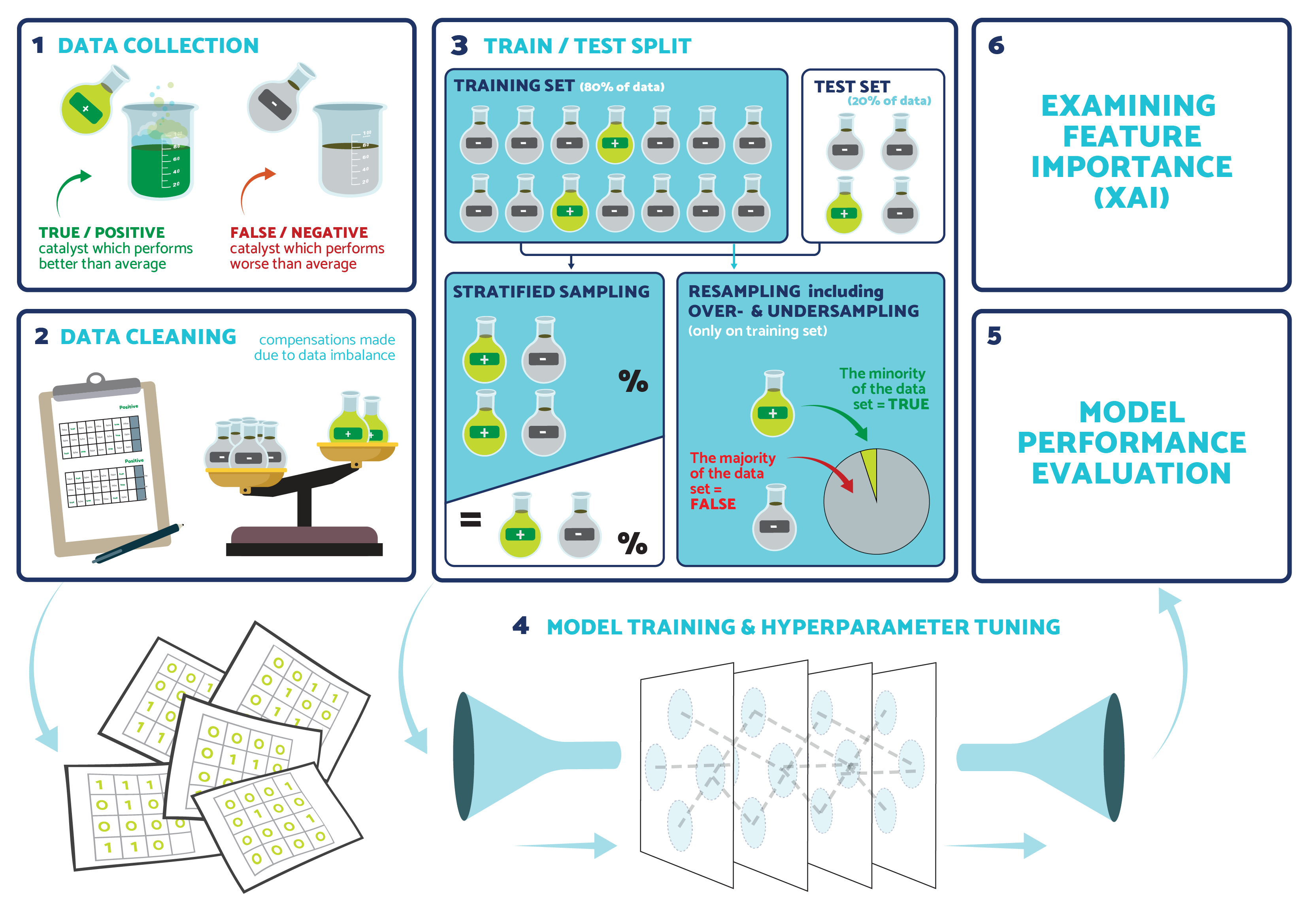}
    \caption{Illustration of the ML framework, starting with data collection and cleaning (steps 1-2), and visualizing the process for obtaining the performance and explanations of a model on a single random train-test split of the dataset (steps 3-6). These training and evaluation steps are then repeated for 100 different train-test splits and the results are aggregated to produce robust performance estimates and feature importance scores.}
    \label{fig:ML framework.png}
\end{figure}

In this section, we will describe the ML framework tailored to address the challenges posed by limited and unbalanced data, illustrated on Figure~\ref{fig:ML framework.png}. 

We begin with data acquisition, followed by data cleaning and preprocessing steps to refine the dataset for further analysis. During the training process, we use stratified sampling to ensure equal representation of all classes across the training and test sets. This is followed by targeted resampling within the training set, addressing the significant imbalance within our dataset of 291 samples, where only 51 are positive. Considering the 80-20 train-test split, this leads to about 230 training samples, with approximately 41 positives initially. In order to balance the dataset for training, we apply SMOTE~\cite{chawla2002smote} with an oversampling ratio of 0.6 for the minority class. This increases the number of positive samples in the training set after resampling to approximately 60\% of the majority class, resulting in about 115 positives. The following undersampling (ratio 1) maintains the total number of samples intact. SMOTE generates new and unique samples by mixing neighboring samples of the minority class, ensuring each is slightly altered and distinct.

The model training includes $k$-fold cross-validation and hyperparameter tuning to improve the predictive accuracy and generalizability of our ML models. We evaluate each model's performance based on accuracy and F1-score metrics. Furthermore, we assess the importance of different features in the dataset via XAI techniques.

To mitigate potential biases due to limited data (291 data points) and variability in the train-test splits, we perform the random splitting of the train and test set and subsequent steps of resampling and evaluation 100 times (steps 3 to 6). The process results in a nested $k$-fold cross-validation, providing reliable model performance estimates through averaging. This ensures better generalizability of our results under small, imbalanced datasets.

\subsection{Robust performance estimation}

\begin{figure}[htbp]
    \centering
    \captionsetup[subfigure]{aboveskip=5mm, belowskip=5mm}
    \begin{subfigure}{.5\textwidth}
        \centering
        \includegraphics[width=\linewidth]{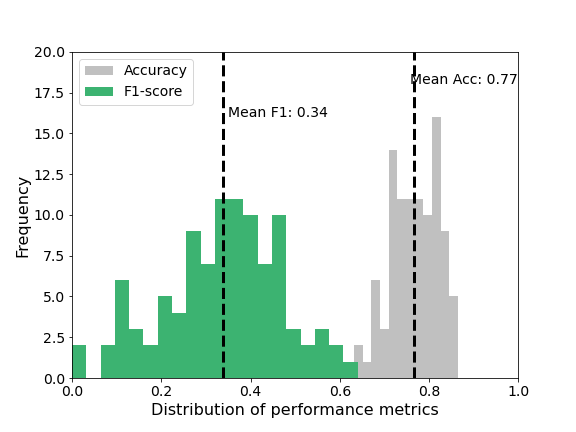}
        \caption{With periodic system group information}
        \label{fig:first_DT_accuracy_distribution}
    \end{subfigure}%
    \begin{subfigure}{.5\textwidth}
        \centering
        \includegraphics[width=\linewidth]{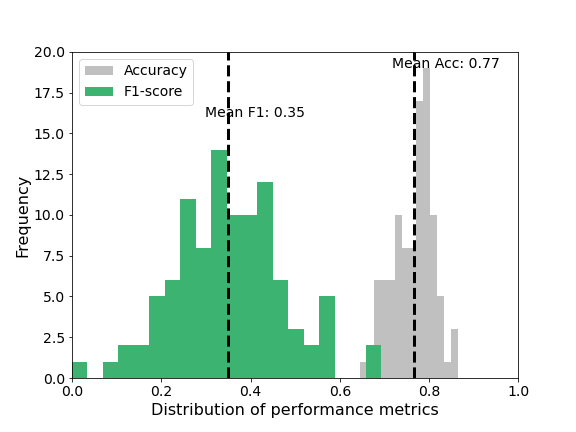}
        \caption{Without periodic system group information}
        \label{fig:pipeline_DT_accuracy_F1_distribution}
    \end{subfigure}
    \begin{subfigure}{.5\textwidth}
        \centering
        \includegraphics[width=\linewidth]{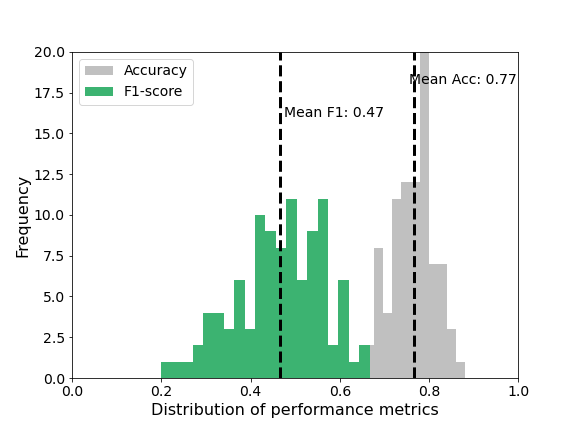}
        \caption{With periodic system group information}
        \label{fig:precision_resampling_withoutgroup}
    \end{subfigure}%
    \begin{subfigure}{.5\textwidth}
        \centering
        \includegraphics[width=\linewidth]{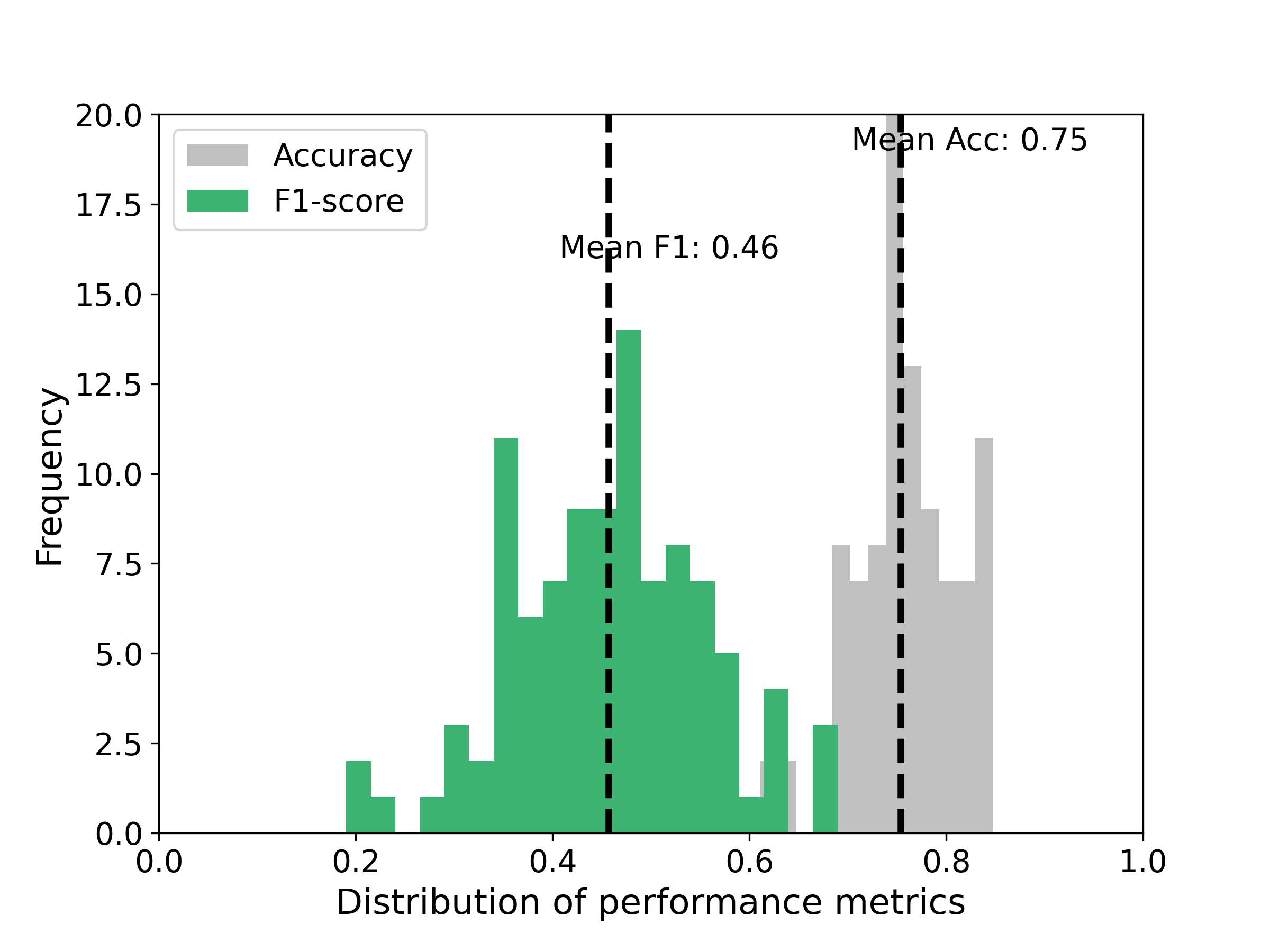}
        \caption{Without periodic system group information}
        \label{fig:pipeline_DT_accuracy_F1_distribution_withoutgroup}
    \end{subfigure}
    \caption{Comparative analysis of accuracy and F1-score distribution for a decision tree over 100 evaluation cycles. Figures (a) and (b) show results without our ML framework, while figures (c) and (d) present results with our ML framework. The vertical line shows the mean of the respective distribution.}
    \label{fig:new Accuracy and F1 distribution}
\end{figure}

In~\citet{nguyen2021learning}, a decision tree model was created using a single train-test split. However, when we tried to replicate this model using an alternative split, we found the model’s performance scores were highly inconsistent. This inconsistency is demonstrated by the wide range of accuracy scores across 100 different splits and random states, as shown in Figure~\ref{fig:new Accuracy and F1 distribution} (a). The accuracy score of the single decision tree model in~\citet{nguyen2021learning} was 0.78, which is very close to the mean of the distribution in the figure, with a value of 0.77.

These variations can be attributed to several factors. First, the randomness inherent in data-splitting results in different subsets being used for training and testing. The sensitivity of decision trees to the training data distribution can, therefore, create inconsistencies in model performance due to these variations in the training set, which are especially high for smaller datasets. In addition, decision tree algorithms often incorporate random initialization of parameters such as feature selection and node splitting thresholds, resulting in different trees being generated at each training iteration, further amplifying the variance of the model's performance.

When dealing with imbalanced datasets, we argue strongly against the reliance on accuracy as the primary performance metric for ML classifiers, due to the susceptibility to misinterpretation of the accuracy scores. When one class significantly outweighs the others, accuracy tends to be skewed, favoring models that simply predict the dominant class. This phenomenon is evident in our dataset, where among 291 data points, only 51 are labeled as positive catalysts. Consequently, if a model consistently labels catalysts as negative, its accuracy would approximate the frequency of the dominant class, yielding a high but misleading accuracy score of 0.82.
Our aim in catalyst material discovery extends beyond recognizing prevalent classes to accurately predicting out-of-distribution samples or classes with fewer samples. The F1-score, by considering both precision and recall and focusing mainly on the positive class, offers a better estimate of a model's performance in imbalanced catalyst design scenarios.

An interesting observation emerges from the comparison between accuracy and F1-score in Figure~\ref{fig:new Accuracy and F1 distribution} (a): while accuracy appears to be satisfactory, F1-score shows a much wider range with lower values. This gap suggests a notable weakness in the model's predictive capacity, particularly for high-yield catalysts.

By using more stable data splitting methods such as stratified sampling and ensuring class balance via resampling strategies, our proposed framework aims to reduce variability and thus improve the reliability of the estimated performance of decision tree-based predictive models. This effect is supported by the fact that the new framework produced a narrower spread of performance metrics, as seen in Figure~\ref{fig:new Accuracy and F1 distribution} (c), indicating a more consistent and robust training process.

The analysis so far has included the features from the periodic system groups introduced in ~\citeauthor{nguyen2021learning} in order to make the comparison fair. However, as we mentioned in Subsection~\emph{\nameref{sec:data}}, we found that these group features do not contribute to the performance of the model. As seen in Figures~\ref{fig:new Accuracy and F1 distribution} (c) and (d), the exclusion of the periodic table group features does not result in any significant change in the distribution of performance scores. Because of this and the issue of explainability as outlined in Supplementary Section~\ref{sup:group_feats}, from here on, we will only report results using the dataset without the periodic table group features, i.e., using only information about which elements and supports were present in the catalyst.
\subsection{Evaluating other machine learning models}
Based on the findings of the previous section, the appropriateness of the decision tree model for our dataset comes under question, prompting us to explore alternative modeling approaches. We first turn our attention to other tree-based techniques such as pre- and post-pruned decision trees, random forest, and XGBoost. To cover a more diverse range of ML approaches, we extended our analysis to non-tree models such as logistic regression, SVMs, and neural networks.

\subsubsection{Performance evaluation}

\begin{figure}[ht]
    \centering
    \includegraphics[width=1\textwidth]{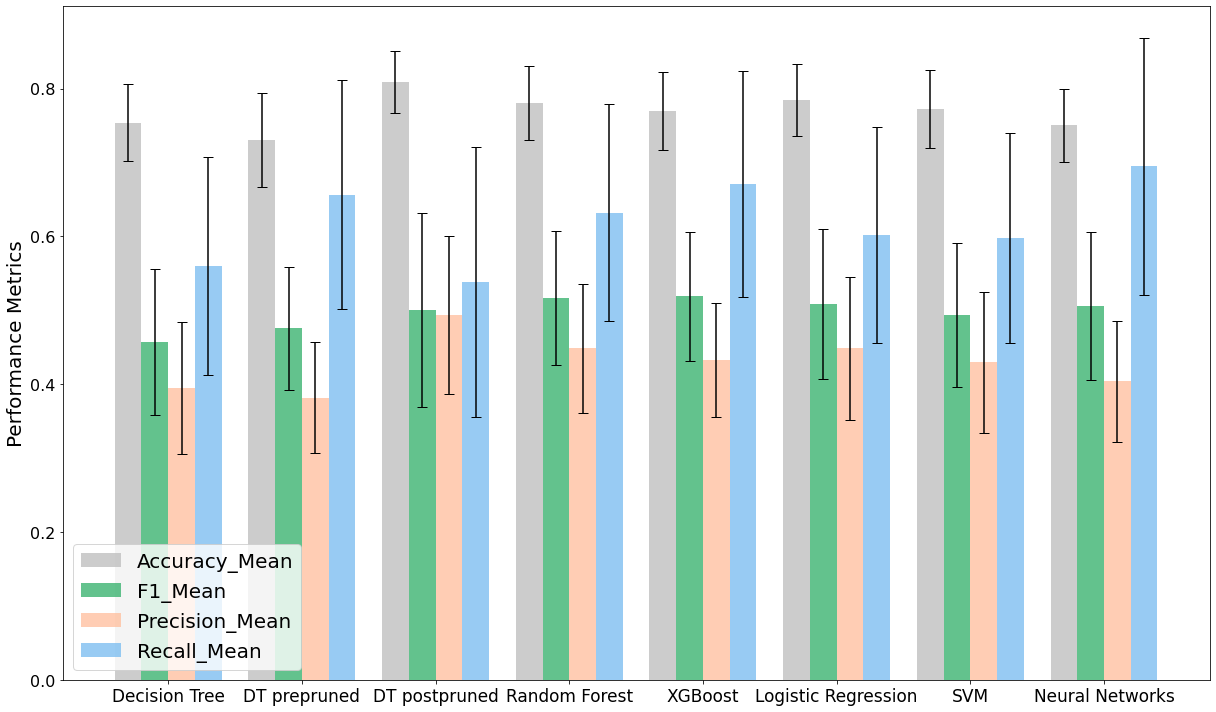}
    \caption{Model performance evaluation with four different evaluation metrics, for all of the discussed ML models. The bars demonstrate the mean of the respective metric, and the error bars present their standard deviation.}
    \label{fig:Model_Performance_Comparision_with_Resampling}
\end{figure}

Following the suggested framework, the performance metrics of all models are calculated and displayed in Figure~\ref{fig:Model_Performance_Comparision_with_Resampling} and Table~\ref{tab:model_evaluation}.
The mean value of accuracy across various models lies between the small range of 0.73 to 0.81. 
The best-performing model is the post-pruned decision tree, with an accuracy of 0.81. However, as we discussed in the previous section, due to the imbalanced ratio of both classes in the OCM dataset, a model classifying all catalyst samples as only having negative performance would have an accuracy of 0.82, demonstrating how misleading using accuracy is as a measure of performance in this case.

A look at the F1-score on the other hand, which takes the under-representation of the high-yield catalysts into account, paints a different picture. For reference, given the dataset's class ratio, the F1-score of a random classifier would be 0.26, a classifier predicting only negative performing catalysts would have an F1-score of 0.0, while a classifier predicting only the positive class would yield an F1-score of 0.3. With this in mind, the results in Table~\ref{tab:model_evaluation} demonstrate that all the models have performed significantly better than the random classifier, with F1-scores ranging from 0.46 to 0.52. Given that this difference in performance was impossible to recognize based on the accuracy, we can conclude that in the context of imbalanced classes, the F1-score is much more informative as a measure of the model's performance compared to accuracy.

\begin{table}[htpb]
\centering
\caption{Model performance evaluation results implemented through the suggested ML framework. The accuracy and F1-score of each model are averaged over 100 training and test splits and compared, and their respective mean and standard deviation are displayed here.}
\label{tab:model_evaluation}
\begin{tabular}{lcccc}
\hline
\textbf{Model} & \textbf{Accuracy Mean} & \textbf{Accuracy Std} & \textbf{F1 Mean} & \textbf{F1 Std} \\
\hline
Decision Tree & 0.75 & 0.05 & 0.46 & 0.10 \\
Decision Tree Prepruned & 0.73 & 0.06 & 0.47 & 0.08 \\
Decision Tree Postpruned & 0.81 & 0.04 & 0.50 & 0.13 \\
Random Forest & 0.78 & 0.05 & \bf{0.52} & 0.09 \\
XGBoost & 0.77 & 0.05 & 0.51 & 0.09 \\
Logistic Regression & 0.78 & 0.05 & 0.51 & 0.10 \\
SVM & 0.77 & 0.05 & 0.49 & 0.09 \\
Neural Networks & 0.76 & 0.05 & 0.51 & 0.10 \\
\hline
\end{tabular}
\end{table}
\FloatBarrier
\subsubsection{Impact of resampling}

\begin{figure}[ht]
    \centering
    \begin{subfigure}{.5\textwidth}
        \centering
        \includegraphics[width=.95\linewidth]{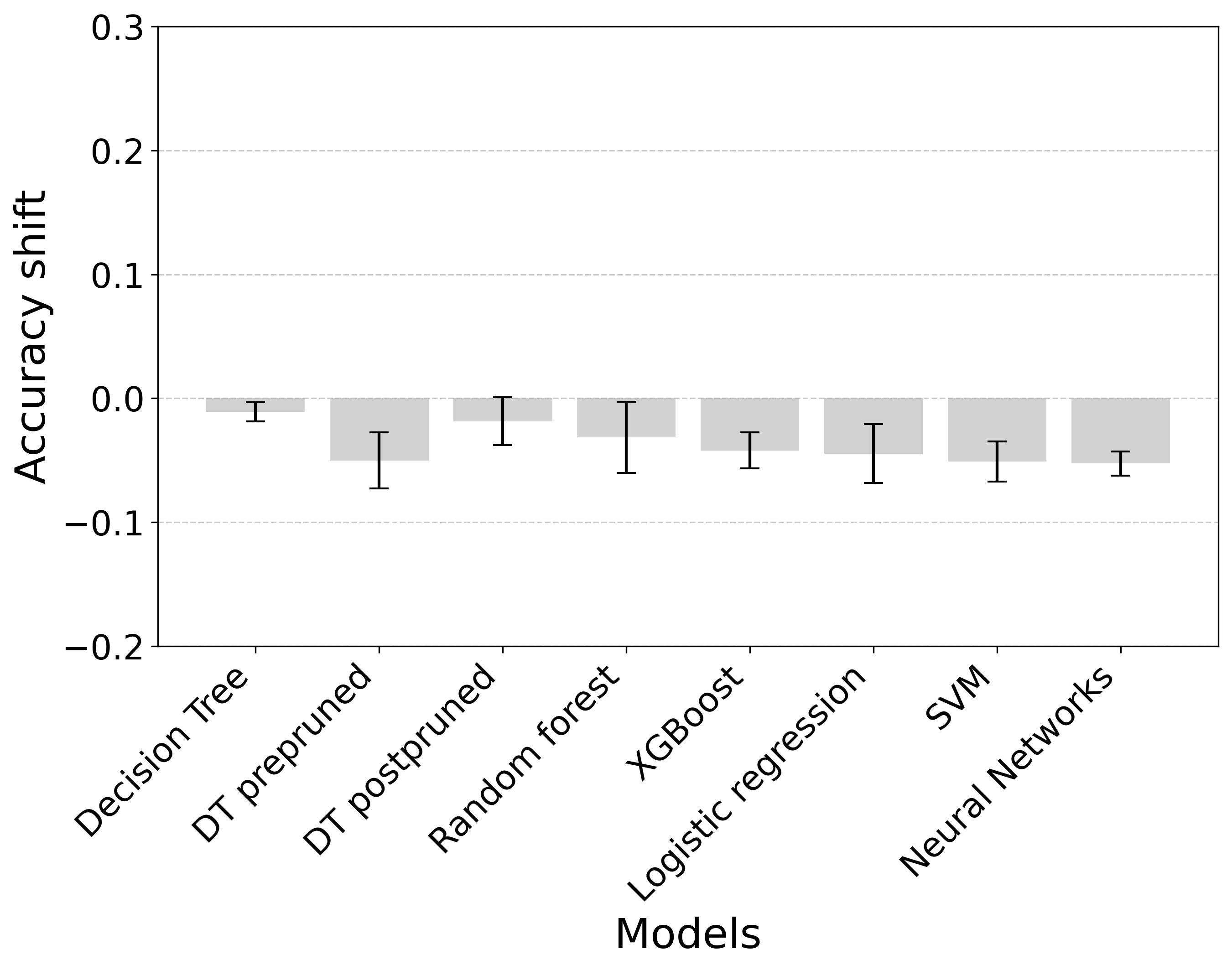}
        \caption{Accuracy}
        \label{fig:accuracy_resampling}
    \end{subfigure}%
    \begin{subfigure}{.5\textwidth}
        \centering
        \includegraphics[width=.95\linewidth]{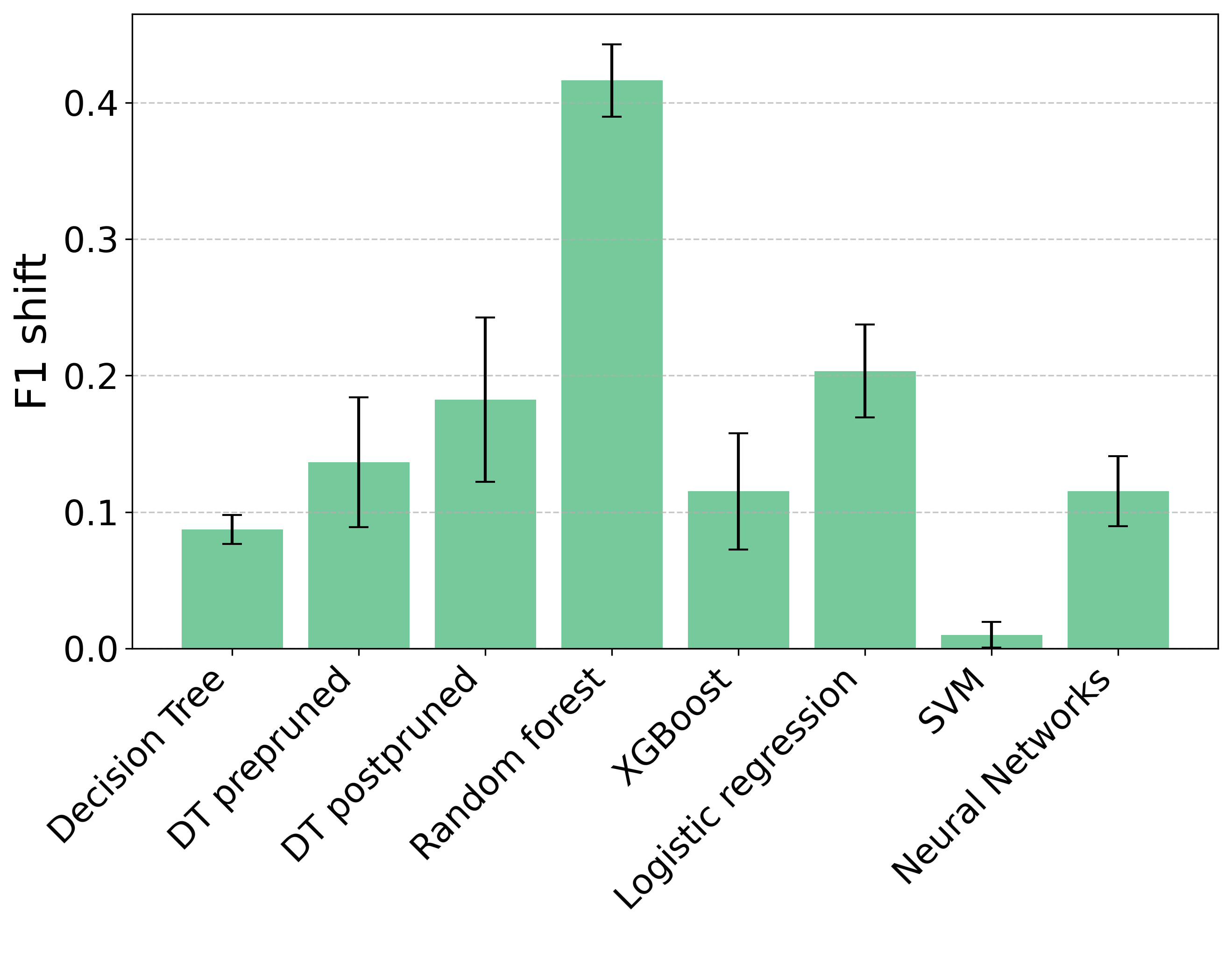}
        \caption{F1-score}
        \label{fig:f1_resampling}
    \end{subfigure}
    \begin{subfigure}{.5\textwidth}
        \centering
        \includegraphics[width=.95\linewidth]{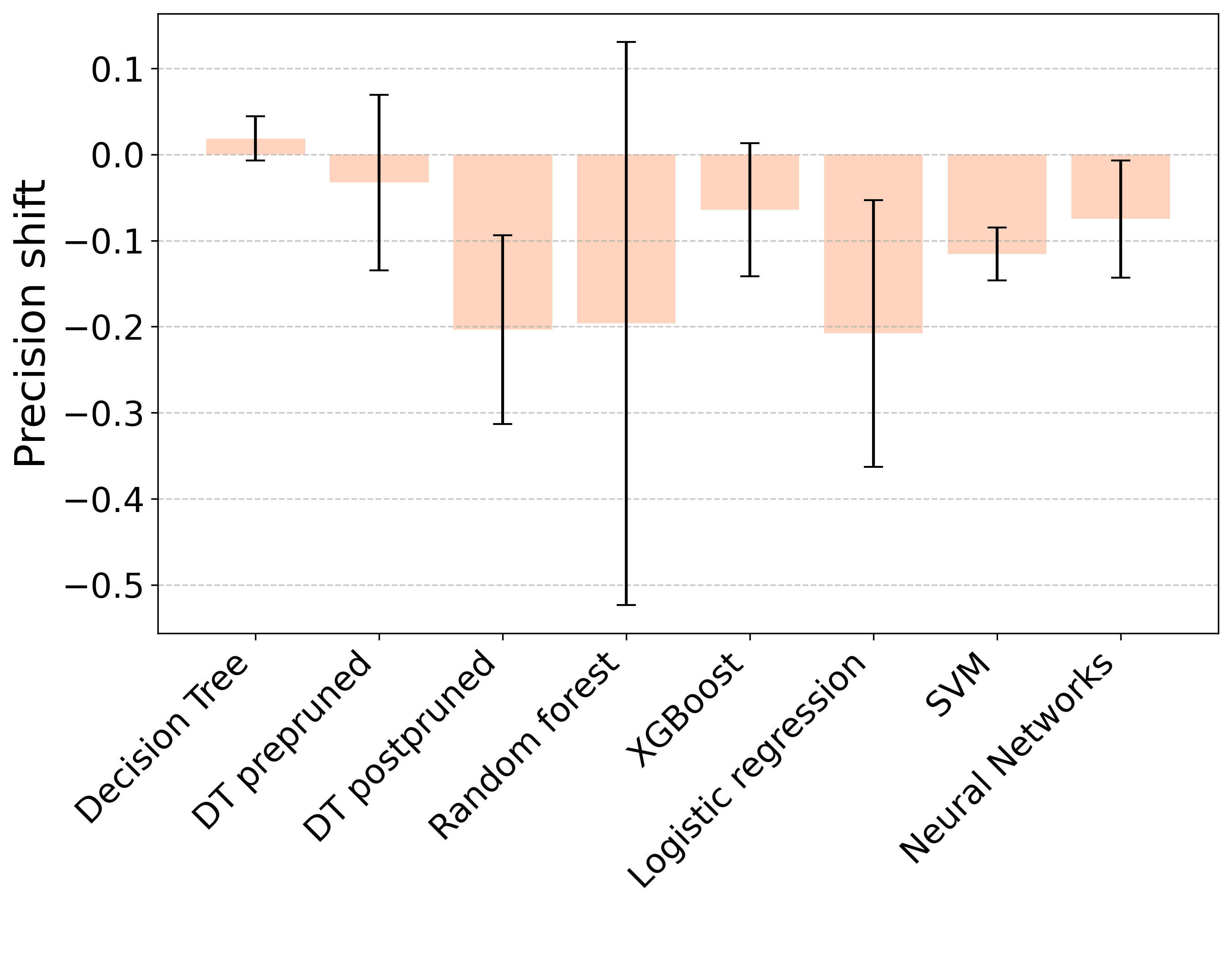}
        \caption{Precision}
        \label{fig:precision_resampling}
    \end{subfigure}%
    \begin{subfigure}{.5\textwidth}
        \centering
        \includegraphics[width=.95\linewidth]{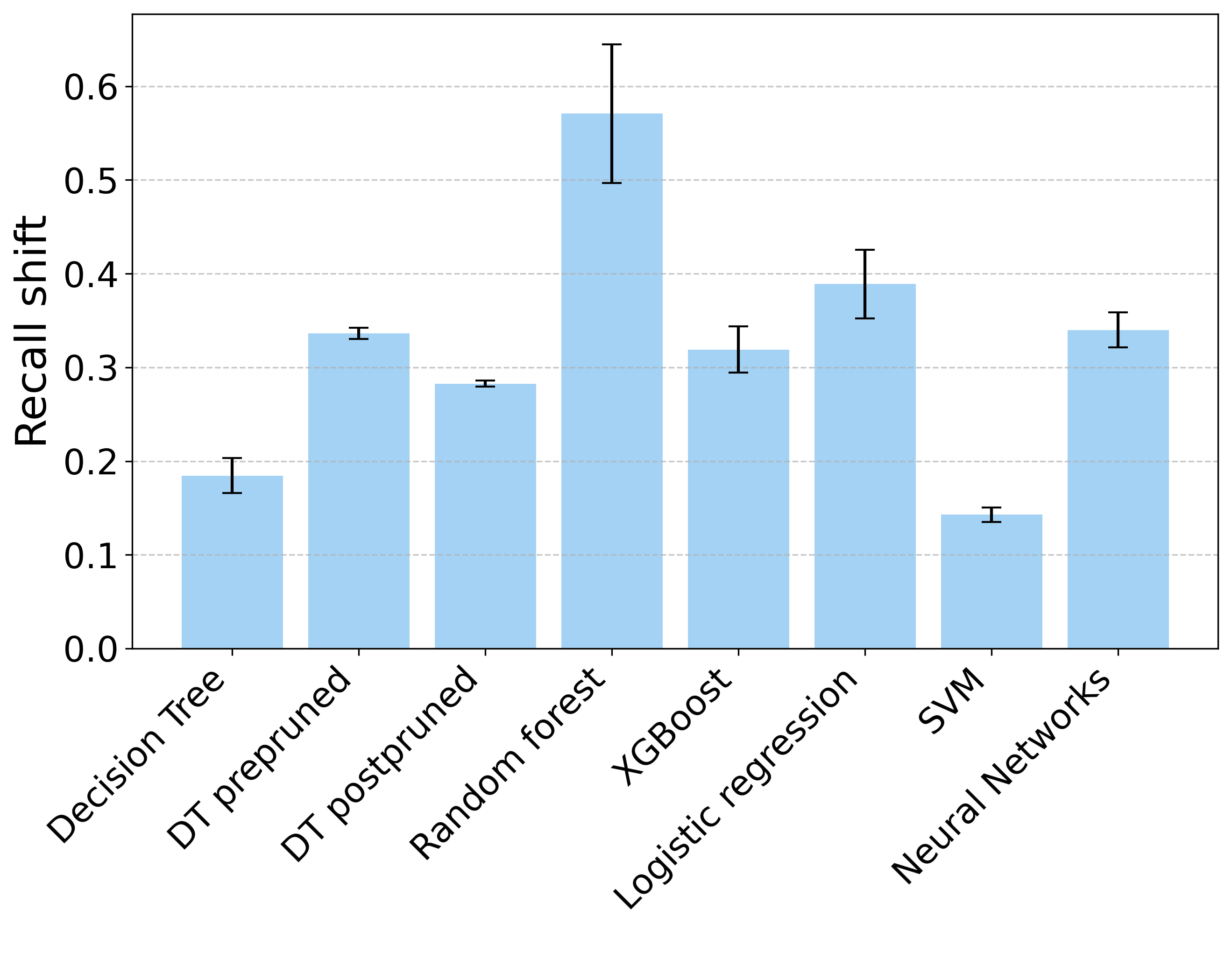}
        \caption{Recall}
        \label{fig:recall_resampling}
    \end{subfigure}
    \caption{The impact of introducing resampling techniques on key performance metrics (a) Accuracy, (b) F1-score, (c) Precision, (d) Recall.}
    \label{fig:resampling_effects}
\end{figure}

To emphasize the benefits of our suggested ML framework, we performed the same evaluation procedure without the resampling step when preparing the training data. Figure~\ref{fig:resampling_effects} illustrates the impact that resampling has on the different performance metrics across all models. We observe a minor drop in accuracy for all models, however, this is of little importance since we already determined that accuracy is not an appropriate performance measure in this context.

\vspace{-1em}

On the other hand, the application of resampling during training has significantly improved the F1-score for all models besides the SVM. The SVM in general is not strongly affected by class imbalance since it relies only on a few samples as support vectors, which usually lie on the edges of the class distributions. A resampling method such as SMOTE that generates artificial samples by mixing existing data points of the same class is thus unlikely to generate any new samples on the edge of the distribution. On the other hand, the random forest has benefited the most from the introduction of resampling, with its F1-score increasing from 0.1 to 0.52.

The overall improvement in F1-score can be attributed to a significant increase in recall across all models. This indicates that resampling enables the models to better identify the minority class of high-yield catalysts, which is the primary class of interest in catalyst design. We also observe a small reduction in precision for most of the models, revealing that the proportion of false positives has slightly increased as a consequence of the models classifying more catalyst compositions as high-yield.

Given the substantial improvements in recall and F1-score, we can confidently conclude that our machine learning framework effectively enhances model performance and reliability for catalyst yield classification.

\subsection{Explaining the decisions of ML models}
Despite the challenges observed in accurately predicting catalyst yield, ML models offer more than just predicting accuracy; they can serve as valuable tools for analysis. In this section, we use the previously trained ML models to explore the underlying factors that drive catalyst performance. For this purpose, we apply a range of XAI methods to identify the most influential features for classifying a sample as 'good' or 'bad'. For each model class, we conducted an aggregation procedure as described in Subsection~\emph{\nameref{sec:evaluation_framework}}: 
For each of the 100 training-test splits, cross-validation was used to identify optimal hyperparameters. These hyperparameters were then used to train a single model on the combined train and validation set for each specific split. The test dataset was subsequently used to estimate the model's generalization performance and generate sample-specific explanations, if necessary. The relevances assigned to all data points were averaged over the number of splits to reduce the effect of model's bias due to specific subsets chosen for training. These aggregated results help identify common patterns and key contributors to catalyst performance, which can provide chemists with insights that can guide future experimental strategies.

\subsubsection{Feature Importance for tree-based models}
\begin{figure}[htbp]
    \centering
    \includegraphics[width=0.65\textwidth]{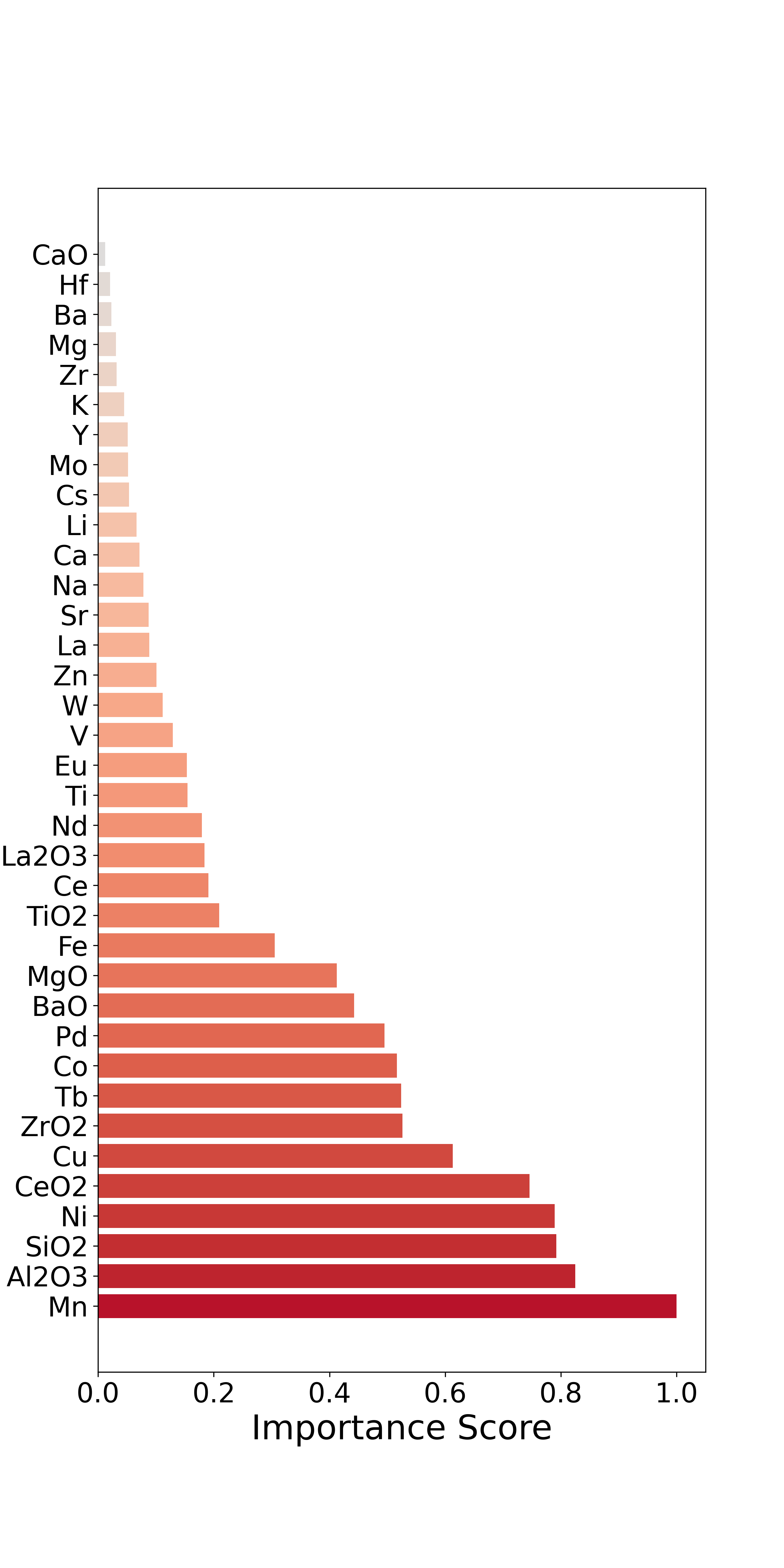}
    \caption{Averaged importance scores for all features across the different tree-based models (Decision tree, DT prepruned, DT postpruned, Random forest, XGBoost)}
    \label{fig:Average_Feature_Importance_Tree_Models}
\end{figure}

In order to aggregate the importance score of features across all tree-based models, we have first normalized these values between zero to one and then took the mean of the importance score for each feature:
\begin{equation}\label{eq:avg_feature_importance_dt}
\bar{R}_d = \frac{1}{S}\sum_{i=0}^S R_d(m^{(i)}),
\end{equation}
where $S$ is the number of training/test splits, $R_d(m^{(i)})$ is the feature importance for feature $d$ extracted from the model $m$ trained on the training subset from split $i$.
The results of the feature importance aggregation are illustrated in Figure~\ref{fig:Average_Feature_Importance_Tree_Models}. Manganese (\ch{Mn}) was found to be the key feature in catalyst yield prediction, followed by Aluminum Oxide (\ch{Al_2O_3}), Silicon Dioxide (\ch{SiO_2}), Nickel (\ch{Ni}) and Cerium Dioxide (\ch{CeO_2}).

Overall, the feature importance scores assigned across the different tree-based models are very similar. As shown in Figure~\ref{fig:Average_corr_coeff_for_all_models}, the correlation coefficients of the importance scores between the tree-based models are all over 0.94. Some part of this high similarity of the feature importances may be explained by the similarity of the explanation methods themselves since the explanations of tree-based models are all based on the reduction of impurity related to each feature. Another and perhaps more significant reason for the similar explanations of the tree-based models is that they all fundamentally use a similar learning strategy of selecting features that reduce the impurity in the leaf/decision nodes.

\subsubsection{Explanations using LRP}\label{sec:lrp_results}

To produce explanations for catalyst yield in SVMs, we performed the neuralization procedure outlined in Section~\emph{\nameref{sec:lrp_for_neuralised_svms}} and applied the propagation rules to obtain relevance scores for each input feature of each test sample. To counteract relevance lost to bias terms, we rescaled input relevance using our rebalancing scheme described in Eq.~\ref{eq:svm_rel_rescaling}.

Similarly to the SVM, neural network explanations are obtained by applying LRP as outlined in Section~\emph{\nameref{sec:lrp}} to each test sample. Again, in order to correct the relevance loss because of the model's bias parameters and maintain the conservation of relevance between the input and output, we rescale the input relevance as shown in Eq.~\ref{eq:nn_rel_rescaling}.

LRP is an explanation method that inherently produces individual explanations for each sample (for some examples of single sample LRP explanations, see Supplementary Section~\emph{\nameref{sec:single_sample_lrp}}). Therefore, to obtain global feature importances based on the entire dataset, it is insufficient to aggregate across the different training/test splits as in Eq.~\ref{eq:avg_feature_importance_dt}, the sample-based explanations within each split also need to be aggregated. This aggregation procedure remains the same for the LRP explanations of both the SVM and neural network models.
The rescaled feature relevances for all test samples are averaged across each sample from each of the 100 test splits:
\begin{equation}\label{eq:avg_feature_importance_lrp}
\bar{R}_d = \frac{1}{S * N}\sum_{i=0}^S\sum_{j=0}^{N} R_d(\mathbf{x}_j^{(i)}),
\end{equation}
where $S$ is the number of training/test splits, $N$ is the number of test samples per split, $R_d(\mathbf{x}_i^{(j)})$ is the relevance for input feature $d$ of the $j$-th sample in the test subset for split $i$.

We stress that LRP explanations yield both positive and negative values, unlike tree-based feature importances, which only produce positive relevance values. Due to our choice of the evidence for the high-yield class as a starting point for the LRP propagation, a positive relevance at the input indicates that this feature contributes positively to the model's prediction of the high-yield class, while the features with negative relevance contribute towards the model classifying the catalyst as low-yield. In contrast, tree-based feature importances only indicate a feature's overall importance without specifying its relation to a particular class.

While aggregated LRP explanations using signed importance scores provide more nuanced information about model behavior, they are not directly comparable to strictly positive tree-based explanations. To enable a direct comparison, we also aggregate absolute LRP relevances across the different samples and splits, providing purely positive feature importances.

\begin{figure}[htbp]
    \centering
    \begin{subfigure}{0.48\textwidth}
      \centering
      \includegraphics[width=\linewidth]{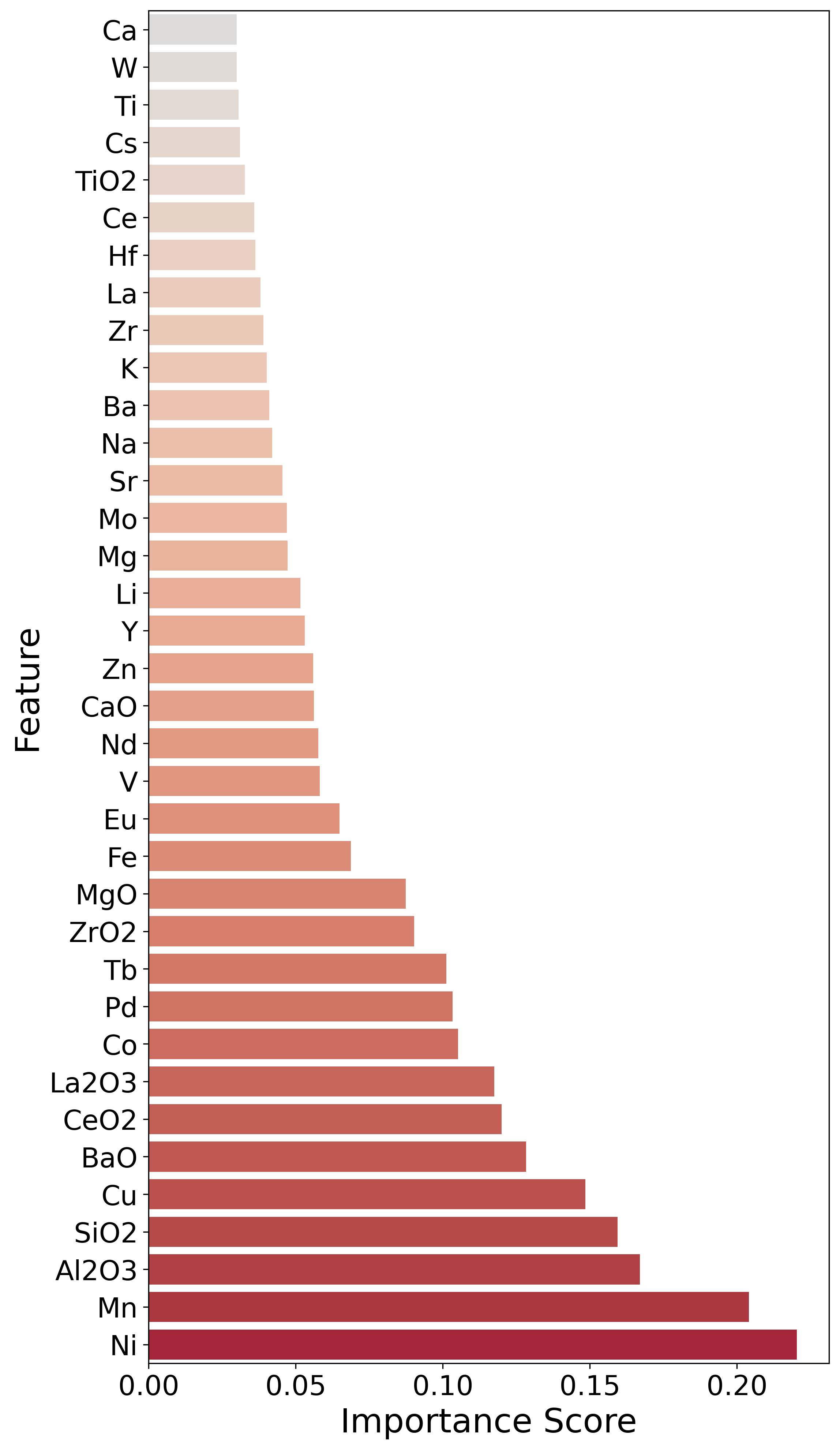}
      \caption{}
      \label{fig:left_nn}
    \end{subfigure}%
    \vspace{0.1cm}
    \begin{subfigure}{0.48\textwidth}
        \centering
        \includegraphics[width=\linewidth]{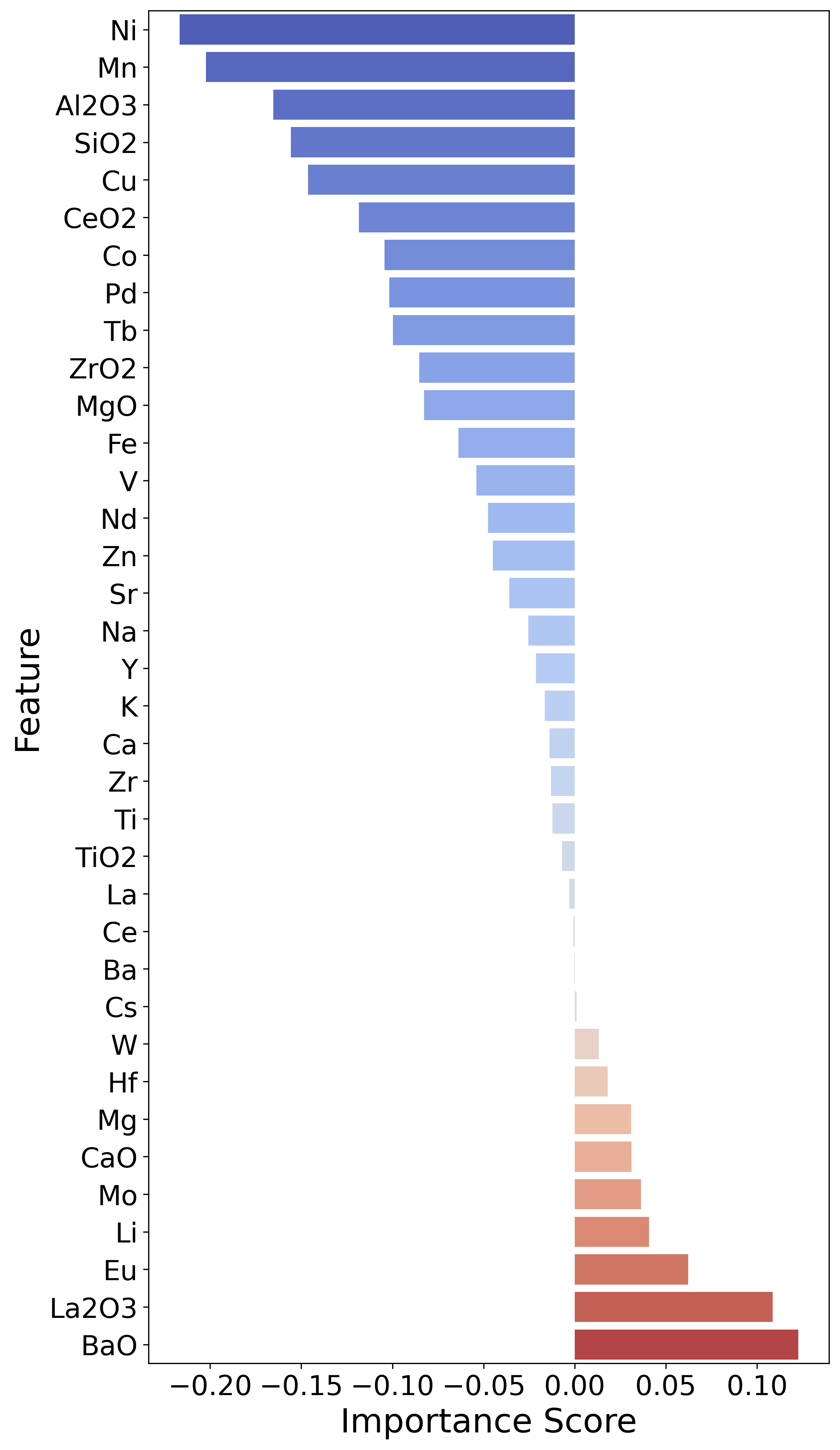}
        \caption{}
        \label{fig:right_nn}
    \end{subfigure} 
    \caption{The mean of feature importance analysis for neural networks via LRP based on the classifier score of the high-yield class - (a): Mean absolute feature importances to identify the key features independently of class-specific relevance. (b): Mean feature relevance, including positive and negative relevances, disentangling the class-specific contributions of the inputs}
    \label{fig:LRP_4_NN}
\end{figure}

\begin{figure}[htbp]
    \centering
    \begin{subfigure}{0.48\textwidth}
        \centering
        \includegraphics[width=\linewidth]{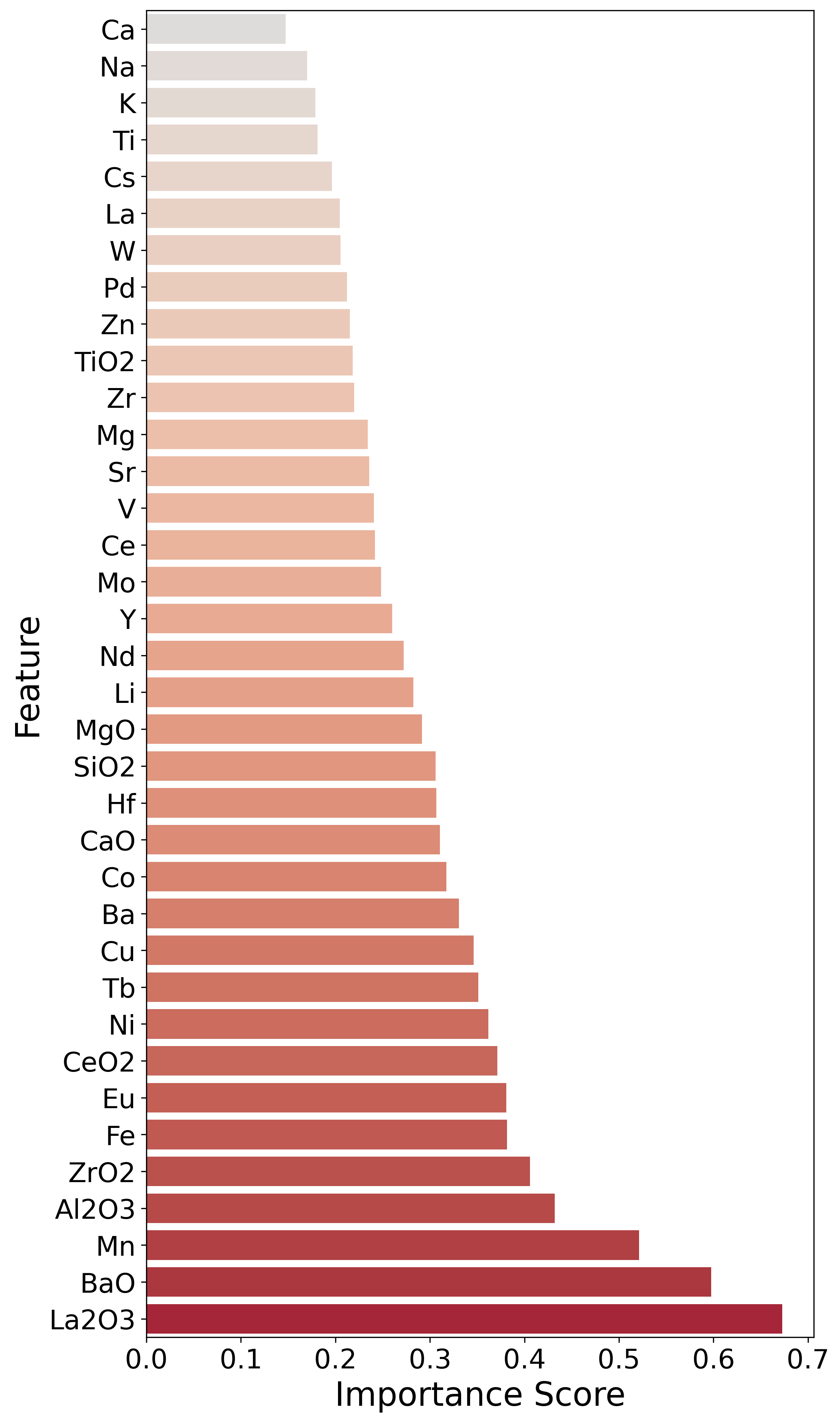}
        \caption{}
        \label{fig:left_svm}
    \end{subfigure}
    \vspace{0.1cm}
    \begin{subfigure}{0.48\textwidth}
        \centering
        \includegraphics[width=\linewidth]{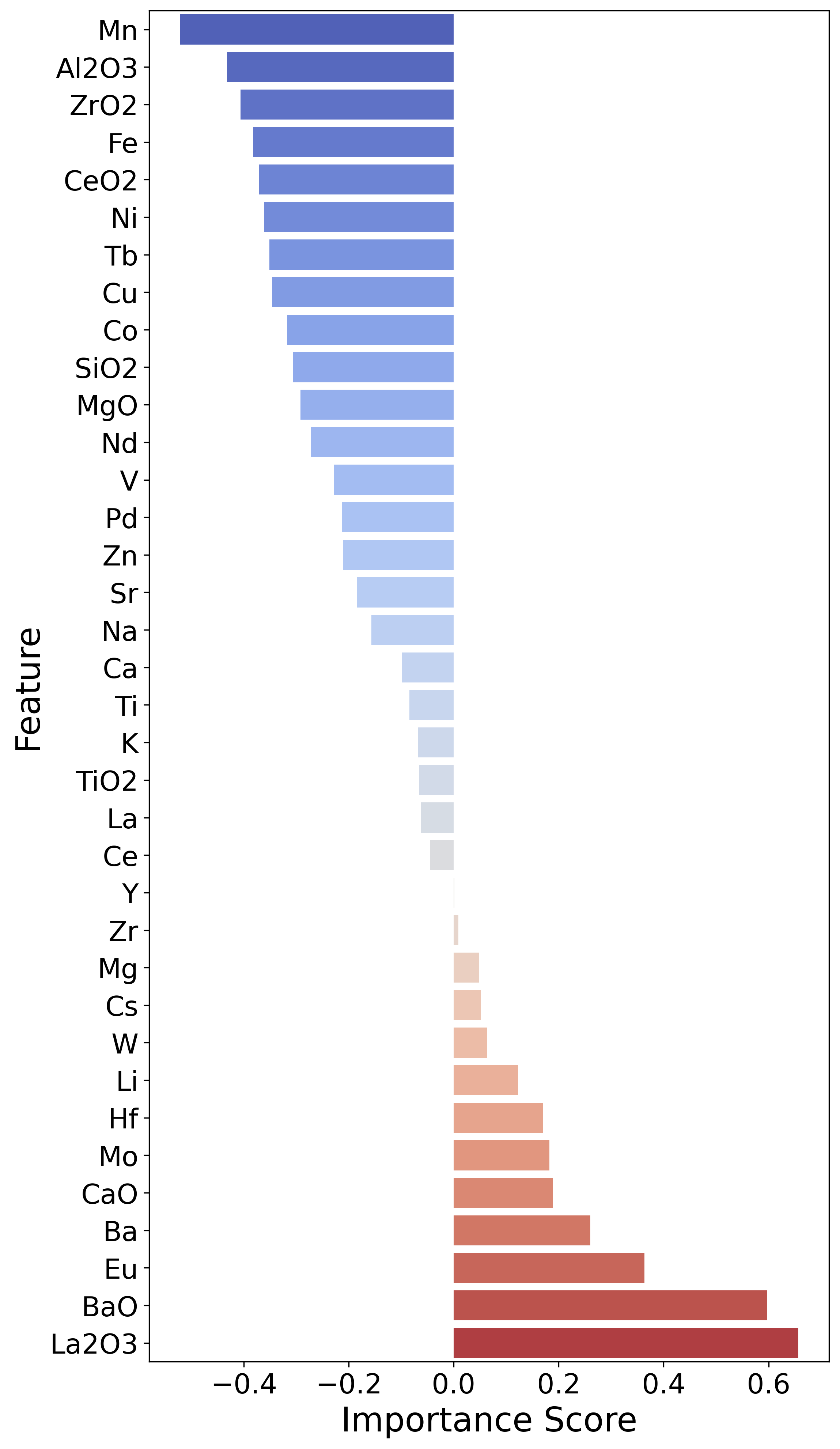}
        \caption{}
        \label{fig:right_svm}
    \end{subfigure} 
    \caption{Feature importance analysis for SVMs via LRP based on evidence for the high-yield class - (a): Mean absolute feature importances to identify the key features independently of class-specific relevance (b): Mean feature relevance including positive and negative relevances, disentangling the class-specific contributions of the inputs.}
    \label{fig:LRP_4_SVM}
\end{figure}

The resulting aggregated signed and absolute feature importances for the SVM model can be seen in Figure~\ref{fig:LRP_4_SVM}, while the analogous visualizations of the average feature importances for the neural network model are shown in Figure~\ref{fig:LRP_4_NN}.

For the neural network models, the highest absolute relevance scores have been assigned to Nickel (\ch{Ni}) and Manganese (\ch{Mn}) alongside alumina (\ch{Al_2O_3}) and silica (\ch{SiO_2}). These components are, therefore, key features for the classification of a catalyst as either high- or low-yield, according to the neural network.
\ch{Mn} and \ch{Al_2O_3} have also been identified as top features by SVM models. However, they are preceded by the supports \ch{La_2O_3}, \ch{BaO}, which have been assigned even higher importance scores.

Thanks to the property of LRP to assign positive and negative relevances to features, the signed averaged LRP importances provide a further dimension for analysis compared to the absolute feature importances. We observe that all of the top absolute contributors identified by both neural networks and SVM have been the highest "negative" contributors to classifying a catalyst as "high-yield", namely \ch{Ni}, \ch{Mn}, and \ch{Al_2O_3}, 
while \ch{La_2O_3}, \ch{BaO} and \ch{Eu} are the key components for classifying a catalyst as high-yield according to both the neural network and SVM models.

\FloatBarrier
\subsubsection{Similarity of explanations}

\begin{figure}[htpb]
    \centering
    \includegraphics[width=0.55\textwidth]{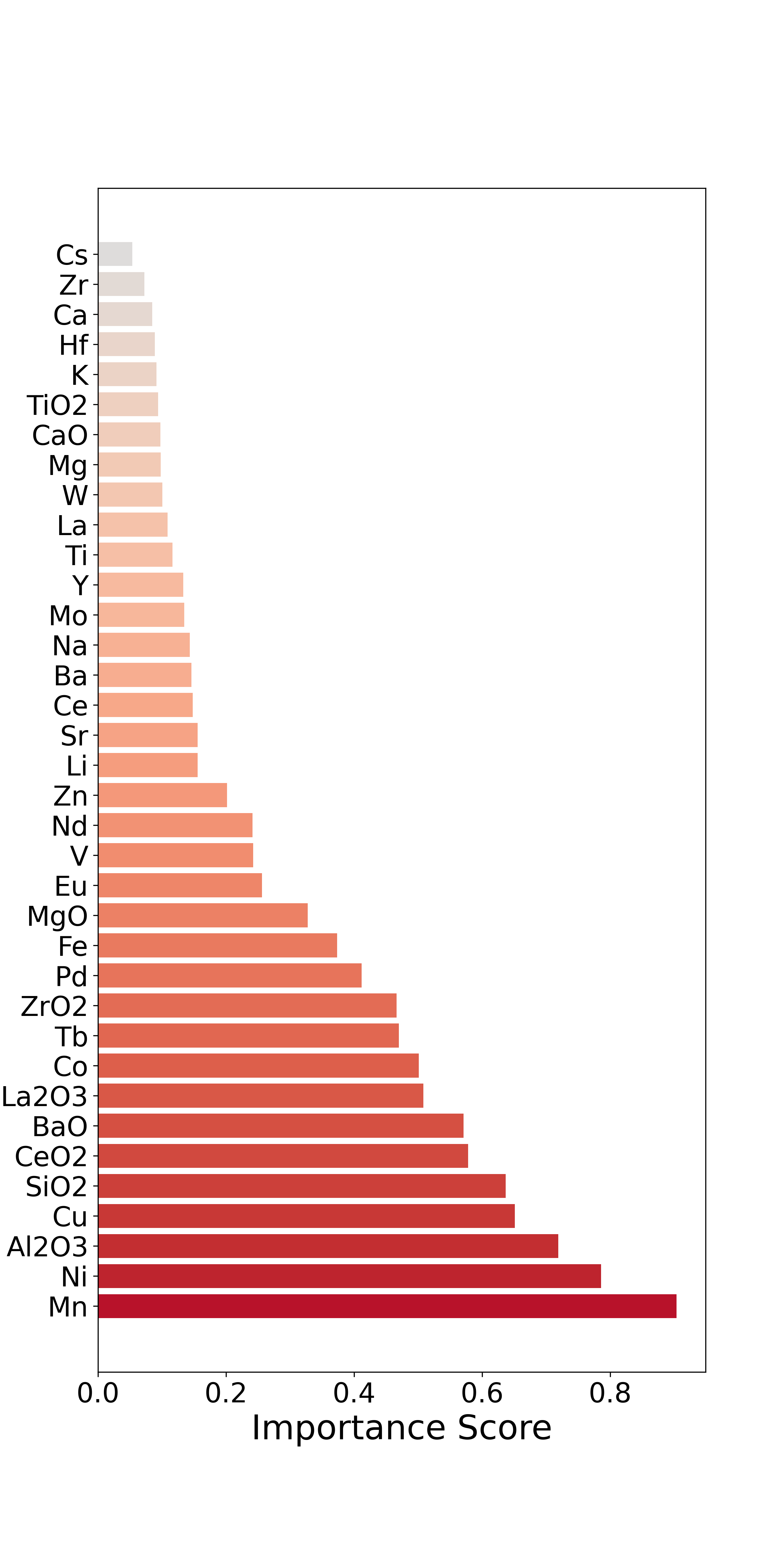}
    \caption{Average feature importance between the models (SVM, Neural networks, Logistic regression, and one tree-based model (Random forest))}
    \label{fig:Average_Feature_Importance_Selected_Models}
\end{figure}

The first thing to note about the absolute feature importances across the tree-based models, neural networks, and SVMs (Figure~\ref{fig:Average_Feature_Importance_Tree_Models}, ~\ref{fig:LRP_4_NN}a and~\ref{fig:LRP_4_SVM}a) is that Manganese (\ch{Mn}) has been identified as one of the most critical elements for determining the yield of a catalyst, accompanied by the support material alumina (\ch{Al_2O_3}), which also has universally high importance. Both of these components are the only ones to appear among the top 5 components in terms of absolute relevance across all three models.

Figure~\ref{fig:Average_Feature_Importance_Selected_Models} visualizes the average absolute feature importance across ML models of different types: SVM, neural networks, logistic regression, and random forest, as a representative of the tree-based models. We find that the top three key metals in determining the yield of a catalyst are Manganese (\ch{Mn}), Nickel (\ch{Ni}), and Copper (\ch{Cu}) and the top three support materials are alumina (\ch{Al2O3}), silica (\ch{SiO_2}) and cerium dioxide (\ch{Ce_2O}).

\begin{figure}[ht]
    \centering
    \includegraphics[width=0.8\textwidth]{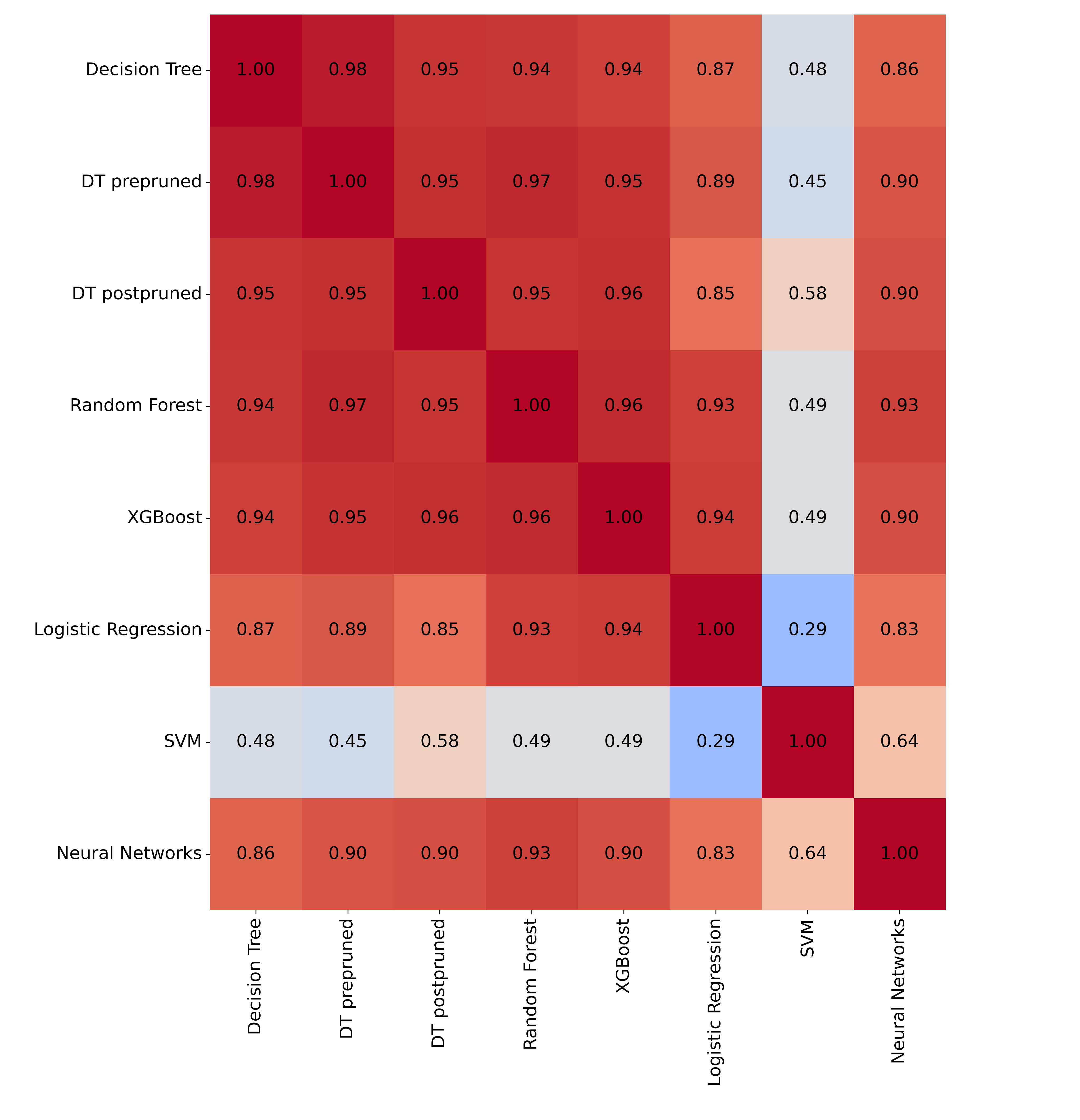}
    \caption{Pearson correlation of the feature importances between all model pairs}
    \label{fig:Average_corr_coeff_for_all_models}
\end{figure}

Furthermore, we performed an analysis aimed at identifying similarities and distinctions between the different models in terms of feature importance. This was achieved by calculating the Pearson correlation coefficients between the feature importance scores of each pair of models via the Fisher-Z transformation (for more details, refer to Supplementary Section~\emph{\nameref{sec:Fisher_Z}}).

The results of this analysis are illustrated through the correlation matrix in Figure~\ref{fig:Average_corr_coeff_for_all_models}, showing us that the feature importance scores of most models are similar to one another. This consistency among the different models and explanation methods indicates that our evaluation framework produces reliable explanations that reflect some underlying phenomena found in the dataset. We also note that for SVMs and neural networks, the correlation analysis was performed using the absolute feature importances to make them directly comparable to the importance scores of the other models (Figures~\ref{fig:LRP_4_NN}a and~\ref{fig:LRP_4_SVM}a). The correlation between the signed feature importances for the SVM and neural network models is 0.90, which is significantly higher than the absolute feature importance correlation of 0.64. 

Additionally, we observe that the SVM model's importance scores display the lowest similarity to those of the other ML models.
Besides differences in applied XAI methods, a likely cause for the low correlation is due to the nature of RBF-kernel models, which by design are unable to perform feature selection and feature weighting. Instead, the total feature relevance is only determined by the choice and distance of local support vectors. Consequently, absolute feature relevance is more uniformly distributed than for other model classes.

\subsection{Discussion of component contributions for high-yield catalyst design}

The analysis in the previous sections suggests that even if explanation methods produce only positive importance scores irrespective of class, it still does not necessarily follow that a component assigned with high importance is beneficial for creating high-yield catalysts. In fact, the component may be deemed important for classification not because it leads to the high-yield catalyst, but because its presence is likely to indicate a low-yield catalyst.

For example, none of the catalysts in the OCM dataset that included \ch{Ni} as a component achieved a high yield. Similarly, only one catalyst containing \ch{Mn} and one containing \ch{Cu} is labeled as high-yield (out of 39 and 31 samples, respectively). Therefore, \ch{Mn}, \ch{Ni}, and \ch{Cu} are indeed key features for determining the yield of the catalyst within the context of this dataset because their presence makes it very likely for a catalyst to be low-yield. This is also reflected in their high absolute importance scores across all models (Figure~\ref{fig:Average_Feature_Importance_Selected_Models}), but more importantly also in the strongly negative relevance for the signed LRP explanations (see Figures~\ref{fig:LRP_4_NN}b and~\ref{fig:LRP_4_SVM}b).

Despite all the above, these results do not necessarily indicate that manganese (\ch{Mn}) is a poor component for OCM catalysis. The explanations provided do not reveal the absolute truth but rather indicate that a specific feature strongly influences the model's classification of a sample as a low-yield catalyst within this particular dataset.
In previous reviews on OCM~\cite{lunsford1995catalytic, lee1988oxidative}, manganese (\ch{Mn}) has been frequently cited as a favorable component, often in combination with sodium (\ch{Na}) and supported by \ch{SiO_2} or \ch{MgO}. However, the current dataset~\cite{nguyen2021learning} consists of 291 catalyst combinations that have been chosen randomly, and not on the basis of previous knowledge, out of a total of 36540 possible combinations. Considering this, it is very likely that the optimal combination of \ch{Mn} with specific elements might be absent from the dataset.
Although the highly negative relevance of \ch{Mn} and other components does not rule out the use of this component in producing high-yield catalysts, it certainly indicates that the component in question may have antagonistic effects when combined with other random components, making it an unattractive candidate for discovering novel high-yield catalysts.

Based on the signed LRP relevance scores, we identify two groups of the contributors to low-yield catalysts: 1) acidic supports, e.g. alumina (\ch{Al_2O_3}) or zirconia (\ch{ZrO_2}), and 2) Highly oxidizing metal oxides, such as \ch{Pd}, \ch{Cu}, \ch{Ni}, \ch{Fe}, \ch{Co}, \ch{Ce}. The supports in group 1) are shown to have a negative impact, especially when they are not neutralized by strong alkali or alkali-earth additives. We argue that this effect is caused by strong adsorption of the ethylene molecule, which is a Lewis base due to its double bond electron pair. This strong adsorption leads to further oxidation towards carbon oxides.
The highly oxidizing elements in group 2) are capable of activating oxygen to strongly oxidizing species that drive the conversion of methane and/or the \ch{C2} coupling products to carbon oxides, reducing the yield of valuable \ch{C2} products. 

On the other hand, positive importance scores are assigned to oxides (either as promoters or as supports) with a higher degree of alkalinity (\ch{BaO}, \ch{CaO}). This effect may arise from the improvement of ethylene desorption, which hinders its further oxidation.

Another group of elements with positive relevance are rare earth oxides, notably \ch{La} and \ch{Eu}. The catalytic activity of rare earth oxides in OCM reaction has been well documented in the literature~\cite{lee1988oxidative}, with \ch{La_2O_3} being one of the best components, alongside \ch{Sm_2O_3}, \ch{Gd_2O_3} and \ch{Er_2O3}. Prior research~\cite{lee1988oxidative} shows that the Lanthanide group plays a role in activating methane as a methyl radical, which is the first step in the coupling of methane to \ch{C2} products. An exception to this is cerium oxide, which has a negative contribution, because cerium, unlike other rare earths studied here, has a reversible valence of Ce4+/Ce3+, making it more oxidizing. This characteristic likely drives the formation of carbon oxides (total oxidation)~\cite{gorte2010ceria}. Our findings of Lanthanum oxides' positive contribution align with the literature. Unfortunately, due to the random choice of components, the other aforementioned rare earth elements are missing from the current dataset.

Our analysis in this section shows that the feature importance scores assigned by our models can be related to chemical phenomena and thus can be used to guide chemists when designing novel high-yield catalysts.

\FloatBarrier
\subsection{Predicting promising catalyst compositions via relevance scores}\label{sec:relevance_sampling}
To demonstrate how feature importances can be used to generate new promising catalyst compositions, we have devised a simple generative algorithm that uses the relevances to bias the generation procedure towards catalysts that the model predicts to be high-yield.

Our algorithm is based on the procedure used to generate the dataset in~\citet{nguyen2021learning}, and ensures that every sample generated is valid, i.e. it could also be generated by their random sampling procedure.

Let $\bar{R}_d$ denote the average relevance for feature $d$ calculated over a given dataset as described in Equations~\ref{eq:avg_feature_importance_dt} or~\ref{eq:avg_feature_importance_lrp}, depending on the type of model and explanation method used. Using this average relevance as an input, the generative algorithm first splits the importance scores into one set for elements and one for supports, after which the two sets of importance scores are converted into discrete probabilities using the softmax function:
\begin{equation}
\mathrm{softmax}(\mathbf{x}, \beta)_i = \frac{e^{\beta \mathbf{x}_i}}{\sum_i e^{\beta \mathbf{x}_i}},
\end{equation}
where $\beta$ is a temperature parameter that can be used to control the variability of the generated samples: a lower value for $\beta$ will result in a more uniform distribution, while a higher value will produce a distribution where the most of the probability concentrated the few components with highest relevances scores. We also utilize two separate temperature parameters, where $\beta^\mathcal{E}$ is used for generating the probability distribution of the elements, while $\beta^{\mathcal{S}}$ is used for the supports.

Based on the probability distributions obtained via softmax, we first sample one support, then sample up to three elements without repetition, each time removing the last sampled element and recalculating the probabilities. Each time an element is sampled, we also include a $\frac{1}{|\mathcal{E}|}$ chance of selecting no elements, which allows our model to sample catalysts with two and one components at the same rate as the sampling method in~\citet{nguyen2021learning}. A detailed step-by-step description can be found in Algorithm~\ref{alg:relevance_sampling}.

\begin{algorithm}
\KwData{\\
$\bar{R}_d$ - set of feature importances\\
$\mathcal{E}$ - feature indices of elements \\
$\mathcal{S}$ - feature indices of supports\\
$\beta^{\mathcal{E}}$ - temperature parameter for the softmax applied on element relevances\\
$\beta^{\mathcal{S}}$ - temperature parameter for the softmax applied on support relevances}
\KwResult{\\
$S^{\mathrm{sel}}$ - feature index of sampled support\\ 
$E^{\mathrm{sel}}_1, E^{\mathrm{sel}}_2, E^{\mathrm{sel}}_3$ - feature indices of sampled elements}
\vspace{10pt}
\nl $\bar{R}^{\mathcal{S}} \gets \{\bar{R}_d: d \in \mathcal{S}\}$;\quad\tcp{Select importances of support features}
\nl $\mathbf{p}^{\mathcal{S}} \gets \mathrm{softmax}(\bar{R}^{\mathcal{S}}, \beta^{\mathcal{S}})$;\quad\tcp{Create probability distribution over supports}
\nl $S^{\mathrm{sel}} \sim \mathbf{p}^{\mathcal{S}}$;\quad\tcp{Sample from the probability distribution over supports}
\nl \For{$i\ \mathbf{\mathrm{in}}\ 1\dots3$}{
    \nl $\bar{R}^{\mathcal{E}} \gets \{\bar{R}_d: d \in \mathcal{E}\}$;\quad\tcp{Select importances of element features}
    \nl $r \sim \mathrm{uniform}(0, 1)$\;
    \nl \eIf(\quad\tcp*[h]{No element is selected with a chance of $1/|\mathcal{E}|$}){$r < \frac{1}{|\mathcal{E}|}$}{
        \nl $E^{\mathrm{sel}}_1 \gets None$\;
    }
    {
        \nl \tcp{Sample element and remove it from the list of indices}
        \nl $\mathbf{p}^{\mathcal{E}} \gets \mathrm{softmax}(\bar{R}^{\mathcal{E}}, \beta^{\mathcal{E}})$\; 
        \nl $E^{\mathrm{sel}}_i \sim \mathbf{p}^{\mathcal{E}}$\;
        \nl $\mathcal{E}.\mathrm{remove}(E^{\mathrm{sel}}_i)$\;
    }
    
  }
\caption{A simple sampling algorithm for generating promising catalyst combinations based on feature importances provided by an explainability method.}\label{alg:relevance_sampling}
\end{algorithm}

\begin{table}[htbp!]
\centering
\caption{Fraction of catalysts generated by our relevance-based sampling procedure that were classified as high-yield by the corresponding ML model. The fractions are reported for the neural network (NN) and XGBoost models, where for the neural network the samples are generated using both the absolute feature importances (NN: abs.) and the signed class-aware feature importances (NN: signed) obtained using LRP. The samples for XGBoost were generated using the absolute feature importances as obtained from the XGBoost model. As the value of the beta parameters increases, we observe that the fraction of samples classified as high-yield decreases when the absolute feature importances are used, while they increase when using class-aware signed feature importances, illustrating the impact of having class-aware importances when using them to guide the development of high-yield catalysts.}
\label{tab:sampling_eval}
\begin{tabular}{@{}lccc@{}}
\toprule
Temperature & \multicolumn{3}{c}{Feature importances}\\
\cline{2-4}
parameters & NN: signed & NN: abs. & XGBoost: abs. \\ \midrule
$\beta^{\mathcal{E}}=10, \beta^{\mathcal{S}}=1$ & 0.38 & 0.17 & 0.31\\
$\beta^{\mathcal{E}}=20, \beta^{\mathcal{S}}=2$ & 0.49 & 0.13 & 0.23\\
$\beta^{\mathcal{E}}=40, \beta^{\mathcal{S}}=4$ & 0.68 & 0.04 & 0.13\\
$\beta^{\mathcal{E}}=40, \beta^{\mathcal{S}}=4$ & 0.85 & 0.01 & 0.05\\
\end{tabular}
\end{table}

Since the feature importances are not the ground truth, but just reflect what the model determines as relevant for prediction, the candidates selected by this procedure are not guaranteed to be high-yield catalysts. However, we can verify the effectiveness of the sampling procedure by feeding the candidates generated with the feature importances back as input into the model that produced these feature importances. If the sampling procedure is effective, then the catalyst candidates produced by this algorithm should be predominately classified as high-yield catalysts.

We performed these experiments for two models with different explanation methods: XGBoost using the impurity metric and neural networks using LRP. For both models, we took the average feature importances (absolute importances for XGBoost and both absolute and signed importances for neural networks) across 100 training/test splits and used them as input into the sampling algorithm to generate 1000 samples with different settings for the $\beta$ parameters.

The results shown in Table~\ref{tab:sampling_eval} confirm our findings from Section~\emph{\nameref{sec:lrp_results}}, about the additional usefulness of having explanations with class-aware feature importances. 

Namely, in the case of the signed feature importances from the neural network, the proportion of generated samples classified as high-yield grows continuously as we use the temperature parameters to bias the sampling more and more towards the high-relevant features. On the other hand, using the absolute feature importances for both the neural network and XGBoost model, we observe that further biasing the sampling distribution towards features with high relevance only produces an increasing number of low-yield catalysts. 

Given that most of the features with high absolute importances are also the ones with highly negative importances (see Figures~\ref{fig:Average_Feature_Importance_Tree_Models} and~\ref{fig:LRP_4_NN}), meaning that they mainly contribute relevance to the class of low-yield catalysts, the results in Table~\ref{tab:sampling_eval} offer further evidence about the reliability of the class-aware LRP explanations.

To summarize, the results in this section indicate that high importance of an element or support in an ML model does not necessarily imply that including this component will produce high-yield catalysts. On the contrary, quite the opposite can be true because a high relevance alone does not give us any information about whether the feature in question predominantly contributes to the desired class. Therefore, drawing conclusions from feature importances requires using explanation methods like LRP, which can disentangle the importance and relationship of a feature to different classes.

\FloatBarrier
\section{Conclusion}

The field of catalyst design is characterized by complex synergistic and antagonistic effects between catalyst components. This often makes high-performing catalysts difficult to discover through traditional trial-and-error methods. Leveraging machine learning's ability to uncover patterns and non-linear relationships, our study aimed to accurately classify the catalytic yield of the OCM reaction based on catalyst composition.
To address the challenges posed by small, unbalanced datasets, we introduced a robust machine learning and XAI framework, incorporating resampling, cross-validation, and well-suited performance measures, as well as XAI techniques that help disentangle the positive and negative contributions of components to catalyst yield.
While we have chosen to apply the framework to OCM as an representative example in this case, the general design of the framework allows it to be applied for various other catalytic reactions.

Our results demonstrated that the accuracy of the various models, both with and without resampling, lies between 76-82\%, which considering the class imbalance in the datasets, is precisely within the range of a random classifier, thereby providing misleading information about model performance.
However, using the F1-score as a performance measure revealed that models with similar accuracy can have significantly different F1-scores (0.1-0.52), allowing for the identification of models who have learned to correctly distinguish the minority class of high-yield catalysts. Having this well-suited performance measure also demonstrated the positive impact of resampling, resulting in an increase of the F1-scores by at least 0.1 across all models, with the random forest model benefiting the most, with an increase of 0.42 in F1-score to reach 0.52. A notable exception to this is the SVM, which by construction is not heavily impacted by class imbalance or resampling.
These findings underscore the effectiveness of our machine learning framework in enhancing model performance and reliability in catalyst yield classification.

The application of various explainable AI techniques consistently identified similar key components influencing models' decisions across different models. Notably, explanations via Layer-wise Relevance Propagation (LRP) effectively disentangled the positive and negative contributions of catalyst components.
Across both SVM and neural networks, LRP explanations have highlighted the same two groups of components are the top positive contributors to high-yield catalysts: rare earth oxides (\ch{La} and \ch{Eu}) and alkaline earth metals with high degree of alkalinity (\ch{Ba}, and \ch{Ca}) as top features in driving high yield catalysts.
These findings, aligning with chemical intuition and existing OCM literature, are notable given the small dataset used. Therefore, future research could focus on applying ML models to larger datasets that encompass diverse catalyst compositions and experiments conducted under various process conditions. This approach could lead to better models and enhanced explanations, contributing towards unveiling previously unknown interactions between components and process conditions.

Despite the limitation of the current dataset, the results presented in this work demonstrate that explainable AI can already be used to extract actionable insights from machine learning models, thereby assisting the chemist in the design of experiments (DOE) for faster discovery of high-yield catalysts.
As a proof of concept, we developed a sampling algorithm based on relevance scores to suggest promising catalyst compositions. The validation of this algorithm using different ML models and XAI methods once again demonstrated the importance of having class-aware relevances for effective catalyst discovery.

Finally, given its unification of robust evaluation practices with interpretable explanations of complex machine learning models, we hope that the ML and XAI framework introduced in this work will provide a useful blueprint for the community, and will promote more reliable and informative analysis of ML models in future work. Our code is available at https://github.com/PSemnani/XAI4CatalyticYield.

\begin{acknowledgement} 

This work was supported by BASLEARN, TU Berlin/BASF Joint Laboratory, co-financed by TU Berlin and BASF SE. 
P.S., M.B., F.B. and K.-R.M. acknowledge support by the German
Federal Ministry of Education and Research (BMBF) for
BIFOLD (BIFOLD24B). K.-R.M. was partly supported by the
Institute of Information \& Communications Technology
Planning \& Evaluation (IITP) grants funded by the government (MSIT) (No. 2019-0-00079, Artificial Intelligence
Graduate School Program, Korea University and No. 2022-0-
00984, Development of Artificial Intelligence Technology for
Personalized Plug-and-Play Explanation and Verification of
Explanation) and by the German Federal Ministry for
Education and Research (BMBF) under Grants 01IS14013BE and 01GQ1115. 
C. W. acknowledges support by BASF Data and AI Academy.
The authors thank Stef Lenk for illustrations (Figures~\ref{fig:abstract_fig.png} to~\ref{fig:ML framework.png}) and also Farnoush Jafari, Laure Ciernik, Rajat Kawade and Jason Hattrick-Simpers for helpful discussions.

\end{acknowledgement}

\FloatBarrier
\clearpage
\begin{suppinfo}
\section{ML models' performance metrics without resampling}\label{sec:ML_no_resampling}

\begin{table}[htbp]
\centering
\caption{Model performance evaluation on accuracy and F1-score without resampling techniques. The accuracy and F1-score of each model are averaged over 100 training and test splits, and their respective mean and standard deviation are displayed.}

\label{tab:data_model_comparison_nr}
\begin{tabular}{lcccc}
\hline
\textbf{Model} & \textbf{Accuracy Mean} & \textbf{Accuracy Std} & \textbf{F1 Mean} & \textbf{F1 Std} \\
\hline
Decision Tree & 0.76 & 0.04 & 0.35 & 0.11 \\
Decision Tree Prepruned & 0.78 & 0.04 & 0.34 & 0.13 \\
Decision Tree Postpruned & 0.82 & 0.02 & 0.32 & 0.19 \\
Random Forest & 0.81 & 0.02 & 0.10 & 0.11 \\
XGBoost & 0.81 & 0.03 & 0.40 & 0.12 \\
Logistic Regression & 0.82 & 0.02 & 0.30 & 0.13 \\
SVM & 0.82 & 0.03 & 0.48 & 0.10 \\
Neural Networks & 0.80 & 0.04 & 0.39 & 0.12 \\
\hline
\end{tabular}
\end{table}

\section{Machine learning models theory}\label{sec:ml_models_theory}

\subsection{Decision Trees}
Decision trees are machine learning models that classify data by iteratively splitting the data along a selected feature at each node~\cite{hastie2009elements}. The feature and value along which the data is split are selected such that they reduce the split's impurity according to some criterion.

A common choice for an impurity measure for classification is the Gini index of a node. Given a node $m$ with a subset $D_m$ of the data of size $N_m$, we define the proportion of class $k$ in the subset as:
\begin{equation}
    \label{eq:pmk}
    \hat{p}_{mk} = \frac{1}{N_m}\sum\limits_{(\mathbf{x}_i, y_i)\in D_m} I(y_i = k),
\end{equation}

where $I$ is the indicator function. We can then define the impurity of node $m$ in terms of Gini index as: 
\begin{equation}\label{eq:gini_index}
\mathcal{L}_{\mathrm{gini}}(m) = \sum_{k \neq k^{\prime}} \hat{p}_{m k} \hat{p}_{m k^{\prime}}=\sum_{k=1}^K \hat{p}_{m k}\left(1-\hat{p}_{m k}\right) = 1 - \sum_{k=1}^K \hat{p}_{m k}^2.
\end{equation}
When deciding how to split a node, the feature and corresponding splitting point are chosen to maximize the information gain, i.e. the reduction of impurity from the parent node $m$ to the split nodes $m_1$ and $m_2$:
\begin{equation}\label{eq:gini_index_o}
IG(m, m_1, m_2) = \mathcal{L}_{\mathrm{gini}}(m) - (\frac{N_{m_1}}{N_{m}}\mathcal{L}_{\mathrm{gini}}(m_1) + \frac{N_{m_2}}{N_{m}}\mathcal{L}_{\mathrm{gini}}(m_2))
\end{equation}

Finally, the data points in every leaf node $m$ are classified as the majority class in the subset $k(m) = \arg\max_k(\hat{p}_{mk})$~\cite{hastie2009elements}.

If the dataset contains a large number of features, the size of the decision tree can be increased until all leaf nodes contain only data points of a single class. While this leads to perfect classification of the training set, it also often results in overfitting when using the decision tree of unseen data, which is why a range of pruning strategies have been developed for controlling the size of the tree. In this study we apply pre-pruning based on the maximum depth of the tree, the minimum number of samples required for splitting, as well as the minimum number of samples in a leaf node. We also use minimal cost-complexity as a post-pruning method~\cite{breiman2017classification}.

\subsection{Random Forest}

Random forest models extend and improve decision trees by using an ensemble of decision tree models~\cite{hastie2009elements}. Each tree model $T_b$ is trained on a bootstrapped subsample of the data complemented by random feature selection. This resulting ensemble reduces the variance of the decision tree models and thereby increases prediction accuracy. The predictions of the random forest classifier are obtained using a plurality vote of the individual decision tree classifiers:
\begin{equation}
    \label{eq:frf}
    \hat{f}^{B}_{rf}(\mathbf{x}) = \arg\max\limits_k \sum_{b=1}^{B} I(T_b(\mathbf{x}) = k),
\end{equation}

where $\hat{f}^{B}_{rf}(\mathbf{x})$ is the prediction of the random forest model for a sample $x$, $B$ is the number of trees, and $T_b(\mathbf{x})$ is the prediction of the $b$-th tree~\cite{hastie2009elements}. 

\subsection{Extreme Gradient Tree Boosting}

Extreme Gradient Tree Boosting (XGBoost) is an ensemble learning method, where weak classifiers in the form of decision tree models are sequentially added to the ensemble in order to minimize the loss~\cite{friedman2001greedy}.
The loss for a given tree structure $q$  at iteration $t$ is given by:

\begin{equation}
\label{eq:XGBeq}
\tilde{L}_t(q) = -\frac{1}{2} \sum_{j=1}^T \underbrace{\left(\frac{(\sum_{i \in I_j} g_i)^2}{\sum_{i \in I_j} h_i + \lambda}\right)}_{\text{Gain of } j\text{-th leaf}} + \gamma T
\end{equation}

For each leave node, the formula computes a scoring metric called gain. The loss operates similarly to the impurity scores in decision trees. In each iteration $t$, we aim to find the loss-minimizing tree structure $q$. Given the impracticality of enumerating all possible tree structures, XGBoost utilizes a greedy algorithm that starts from a single leaf and iteratively adds branches. The algorithm splits a leaf if the summed gain of the left and right nodes after the split is larger than the original node gain and a regularization term $\gamma$, as detailed by Equation~\ref{eq:XGBscore}. 
\begin{equation}
\label{eq:XGBscore}
L_{\text{split}} = \frac{1}{2} \left[ \underbrace{\frac{(\sum_{i \in I_L} g_i)^2}{\sum_{i \in I_L} h_i + \lambda}}_{\text{Gain}_{\text{left}}} + \underbrace{\frac{(\sum_{i \in I_R} g_i)^2}{\sum_{i \in I_R} h_i + \lambda}}_{\text{Gain}_{\text{right}}} - \underbrace{\frac{(\sum_{i \in I} g_i)^2}{\sum_{i \in I} h_i + \lambda}}_{\text{Gain}_{\text{original}}} \right] - \gamma
\end{equation}

The algorithm splits a leaf based on the maximum gain obtained after splitting the node over all possible splits. Splitting is done recursively and constrained by the regularization parameters $\gamma$ and $\lambda$. We refer to the original paper for more details \cite{chen2016xgboost}.

\subsection{Logistic Regression}

Logistic regression models the probability of binary outcomes conditional on input features:

\begin{equation}
    \label{eq:logistic_regression}
    P(y=1|\mathbf{x}) = \frac{1}{1 + e^{-(\theta^T \mathbf{x})}}.
\end{equation}

$P(y=1|\mathbf{x})$ denotes the probability that an sample $\mathbf{x}$ belongs to class 1 conditional on the input features. The coefficients $\theta$ are then optimized in order to minimize the logistic loss: 
\begin{equation}
    \label{eq:loss_function}
    \mathcal{L}(\hat{y}, y) = -y\log \hat{y} - (1 - y)\log \hat{y},
\end{equation}

which in the binary case is proportional to the cross entropy loss~\cite{hastie2009elements}.

\subsection{Support Vector Machine}

The SVM minimizes the empirical risk by finding an optimal hyperplane that maximizes the margin between different classes~\cite{boser1992training, muller2001introduction} using a linear classifier of the form $f(\mathbf{x}) = \mathbf{w}^T\mathbf{x} + b$, where $\mathbf{w}$ is the vector that defines the hyperplane and $b$ is a bias term. This formulation yields the following quadratic optimization problem:
\begin{equation}
    \label{eq:svm}
    \min_{\mathbf{w}, b} \frac{1}{2}||\mathbf{w}||^2 + C\sum\limits_{i=1}^n\xi_i \quad \text{subject to} \quad y_i (\mathbf{w} \cdot \mathbf{x}_i + b) \geq 1, \, \forall i,
\end{equation}
where $\xi_i$ are slack variables introduced to the model that can be applied in cases where the classes are not perfectly separable, and $C$ is a regularization parameter that determines the trade-off between empirical risk and model complexity.

Mapping the features $\mathbf{x}$ to a higher dimensional space using a feature map $\phi(\mathbf{x})$ allows the SVM to perform non-linear classification in the original input space by finding a maximal margin hyperplane in the feature space. By solving the optimization problem in the dual space, we obtain the following decision rule for the SVM classifier:
\begin{align}\label{eq:svm_pred_fun}
f(\mathbf{x}) & =\operatorname{sgn}\left(\sum_{i=1}^n y_i \alpha_i\left(\phi(\mathbf{x})^T\phi\left(\mathbf{x}_i\right)\right)+b\right) \\
& =\operatorname{sgn}\left(\sum_{i=1}^n y_i \alpha_i \mathrm{k}\left(\mathbf{x}, \mathbf{x}_i\right)+b\right),
\end{align}
where $k\left(\mathbf{x}, \mathbf{x}_i\right) = \phi(\mathbf{x})^T\phi(\mathbf{x}_i)$ is the so-called kernel function, the coefficients $\alpha_i$ are called the dual coefficients, and $b$ is a bias term. One of the most commonly used kernels for machine learning, and the one we use in this work is the radial basis function (RBF) kernel, defined as follows:
\begin{equation}
    \label{eq:rbf_kernel}
    k\left(\mathbf{x}, \mathbf{x}^{\prime}\right)=\exp \left(-\gamma \cdot \lVert \mathbf{x - x'}\rVert^2 \right),
\end{equation}

Given that the kernel function satisfies Mercer's conditions~\cite{mercer1909xvi}, we can guarantee that there is an associated feature map $\phi$, so by expressing the optimization problem and decision rule in terms of inner products of feature maps, we can bypass the feature maps and work directly with the kernel functions, allowing us to implicitly operate in high- and even infinite-dimensional features spaces~\cite{muller2001introduction}.

\subsection{Neural Networks - Multi-Layer Perceptron (MLP)}
Multi-layer perceptrons (MLPs) are among the first and simplest neural network architectures to be conceived. 
The map of the input features $\mathbf{x}$ to an output $y$ by transforming the input using a succession of layers. 
Each layer is composed of a linear map consisting of a weight matrix and $W^l\in \mathbb{R}^{d_{l-1}\times d_l}$ and a bias vector $\mathbf{b}^l\in\mathbb{R}^{d_l}$, where $d_l$ and $d_{l-1}$ are the dimensionalities of the layers $l$ and $l-1$, followed by a non-linearity $\sigma^l$. This gives the multi-layer perceptron the overall functional form of:

\begin{equation}
    \label{eq:neural_network}
    f_\theta(\textbf{x}) = \sigma^{L}(W^{L}(\sigma^{L-1}(...\sigma^1(W^1\mathbf{x} + \mathbf{b}^1))) + \mathbf{b}^L),
\end{equation}

where $\theta$ denotes the set of parameters of the network, i.e. the values of the weight matrices and bias vectors. The network parameters are optimized via gradient descent to minimize some loss function between the network's output and target $\mathcal{L}(f_\theta(\textbf{x}), y)$, where the backpropagation algorithm is used to efficiently update the parameters of the network~\cite{rumelhart1986learning}. Due to the large number of parameters and non-linearities, the loss function of the neural network is usually highly non-convex, which means the gradient descent often converges to local optima of the loss function, though this usually does not have a significant effect on the performance of the MLP~\cite{lecun2002efficient,choromanska2015loss,zhang2021understanding}.

\clearpage
\section{ML models training and feature importance scores results}\label{sec:ml_models_results}
In this section, we have provided more details on each of the ML models' training hyperparameters. Additionally, the average importance of each feature is plotted (a), which is a normalized value, with a minimum of 0 and a maximum of 1. Moreover, the distribution of the importance of each feature is illustrated (b), with original values (not normalized).

\FloatBarrier
\subsection{Decision tree models}

\begin{figure}[htbp!]
    \centering
    \begin{subfigure}{0.48\textwidth}
        \centering
        \includegraphics[width=\linewidth]{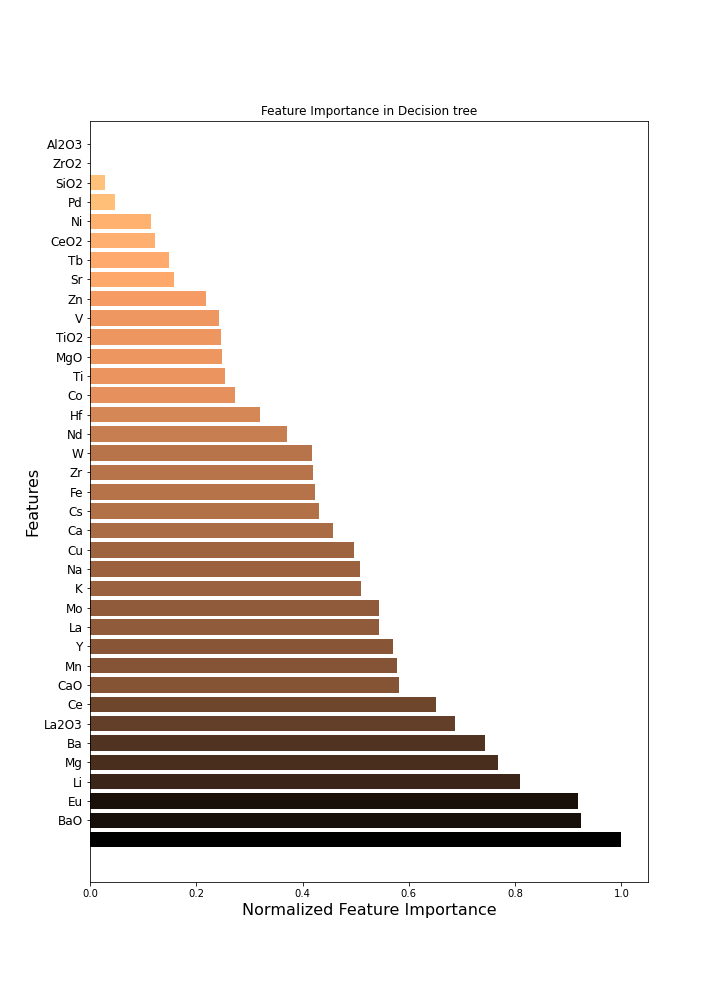}
        \caption{}
        \label{fig:left_dt}
    \end{subfigure}%
    \vspace{0.1cm}
    \begin{subfigure}{0.48\textwidth}
        \centering
        \includegraphics[width=\linewidth]{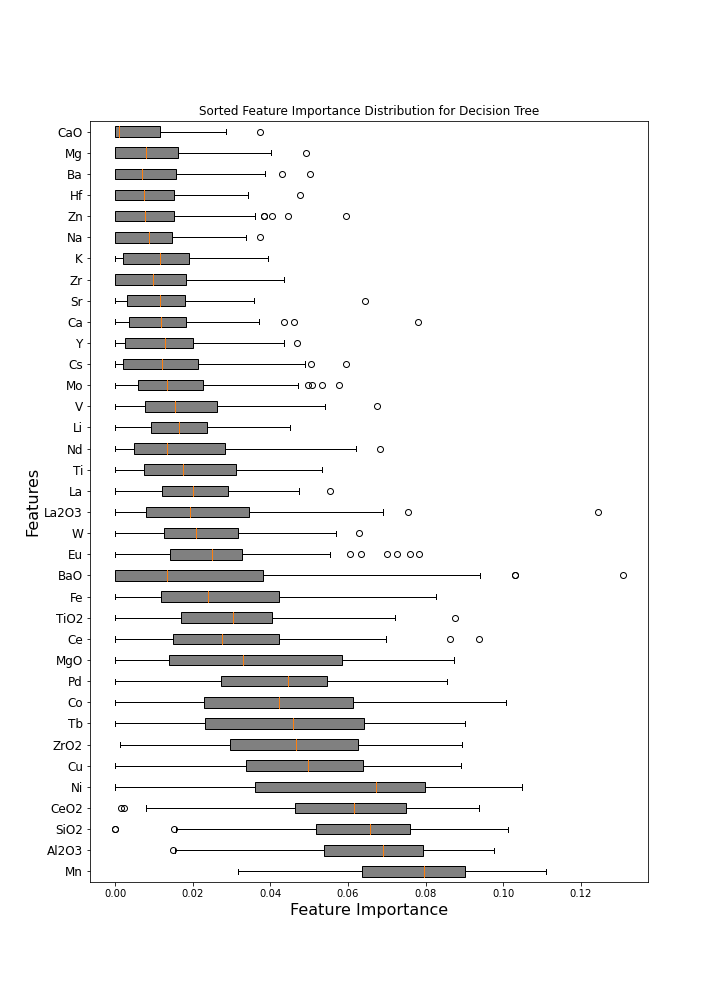}
        \caption{}
        \label{fig:right_dt}
    \end{subfigure}
    \caption{Feature importance analysis for decision tree models}
    \label{fig:FI_4_DT}
\end{figure}

\FloatBarrier
\clearpage

\subsection{Decision tree with prepruning models}

\begin{table}[htbp!]
\centering
\caption{Hyperparameters Tuned in the Decision Tree with prepruning Classifier}
\label{tab:hyperparameters}
\begin{tabular}{@{}lll@{}}
\toprule
Hyperparameter       & Description                                                      & Range      \\ \midrule
\texttt{max\_depth}         & Maximum depth of the tree.                                       & 1--10      \\
\texttt{min\_samples\_split} & Minimum number of samples required to split an internal node.   & 2--20      \\
\texttt{min\_samples\_leaf}  & Minimum number of samples required to be at a leaf node.        & 1--20      \\ \bottomrule
\end{tabular}
\end{table}

\begin{figure}[htbp!]
    \centering
    \begin{subfigure}{0.48\textwidth}
        \centering
        \includegraphics[width=\linewidth]{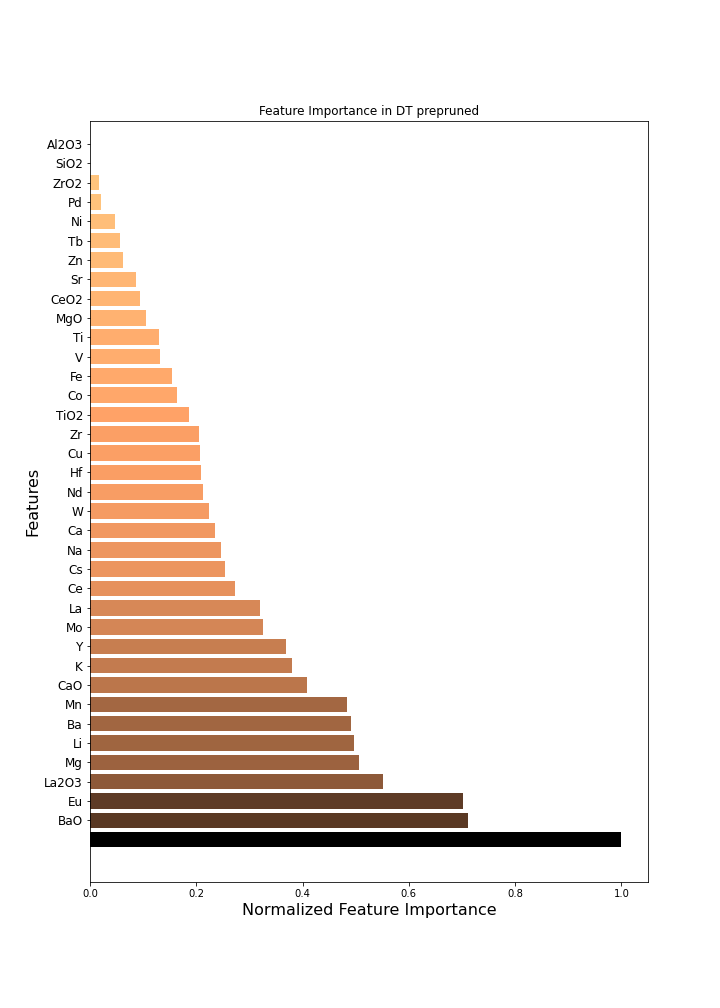}
        \caption{}
        \label{fig:left_dt_pre}
    \end{subfigure}%
    \vspace{0.1cm}
    \begin{subfigure}{0.48\textwidth}
        \centering
        \includegraphics[width=\linewidth]{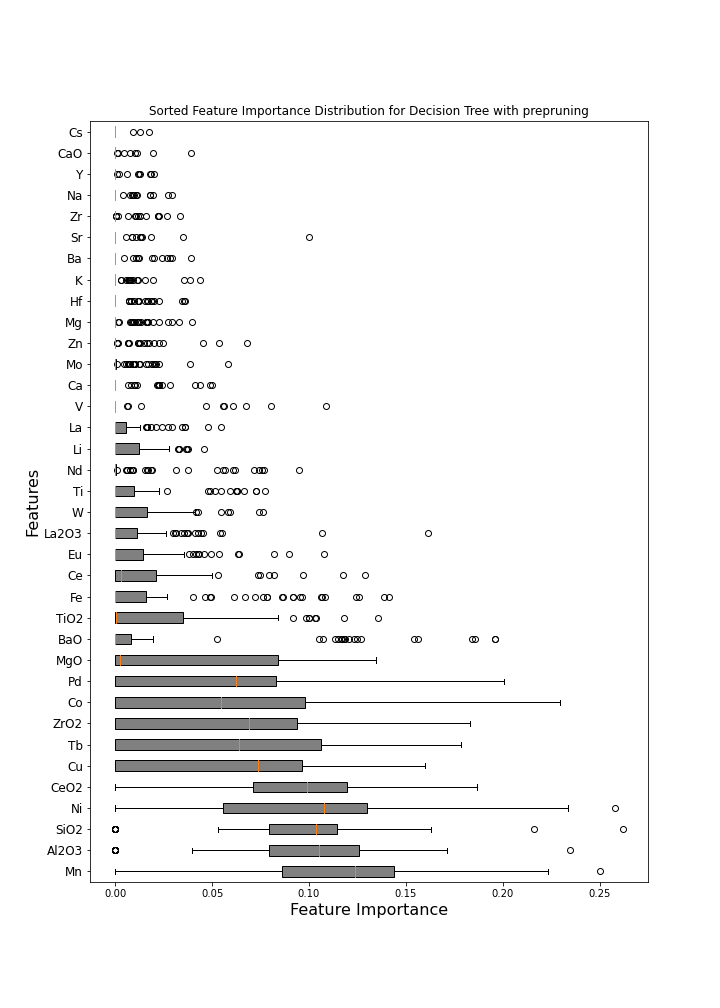}
        \caption{}
        \label{fig:right_dt_pre}
    \end{subfigure}
    \caption{Feature importance analysis for decision tree with prepruning models}
    \label{fig:FI_4_DT_pre}
\end{figure}

\FloatBarrier
\clearpage
\subsection{Decision tree with postpruning models}

\begin{table}[htbp!]
\centering
\caption{Hyperparameters Tuned in the Decision Tree with postpruning models}
\label{tab:hyperparameters_nr}
\begin{tabular}{@{}lll@{}}
\toprule
Hyperparameter       & Description                                                      & Range      \\ \midrule
\texttt{max\_depth}         & Maximum depth of the tree to prevent overfitting.               & 1--10      \\
\texttt{min\_samples\_split} & Minimum number of samples required to split an internal node.   & 2--20      \\
\texttt{min\_samples\_leaf}  & Minimum number of samples required to be at a leaf node.        & 1--20      \\ \bottomrule
\end{tabular}
\end{table}

\begin{figure}[htbp!]
    \centering
    \begin{subfigure}{0.48\textwidth}
        \centering
        \includegraphics[width=\linewidth]{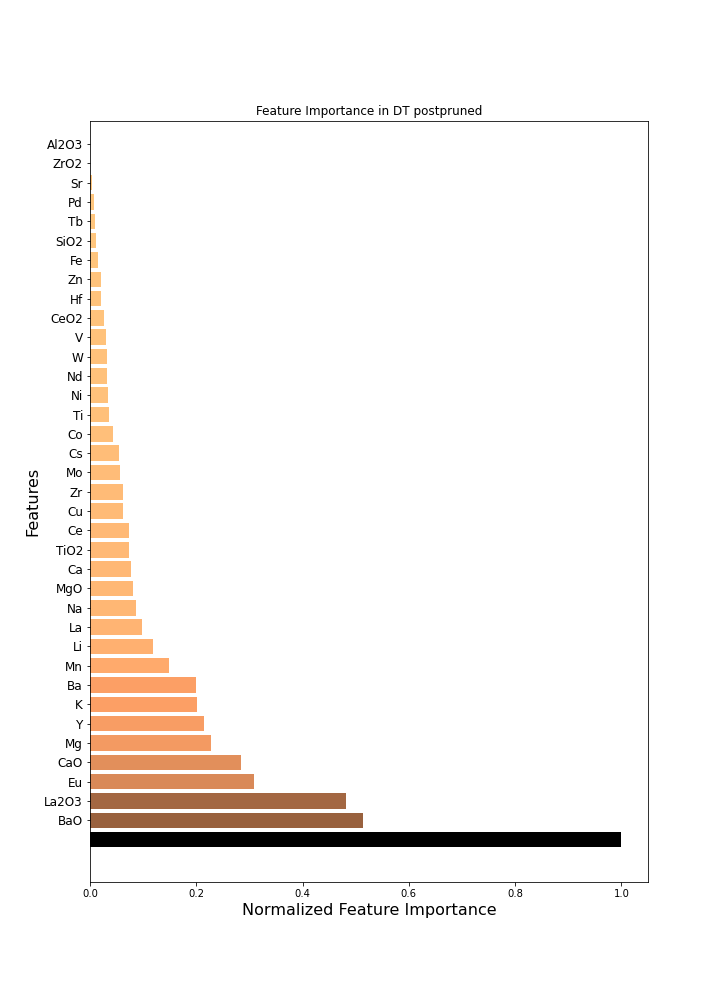}
        \caption{}
        \label{fig:left_dt_post}
    \end{subfigure}%
    \vspace{0.1cm}
    \begin{subfigure}{0.48\textwidth}
        \centering
        \includegraphics[width=\linewidth]{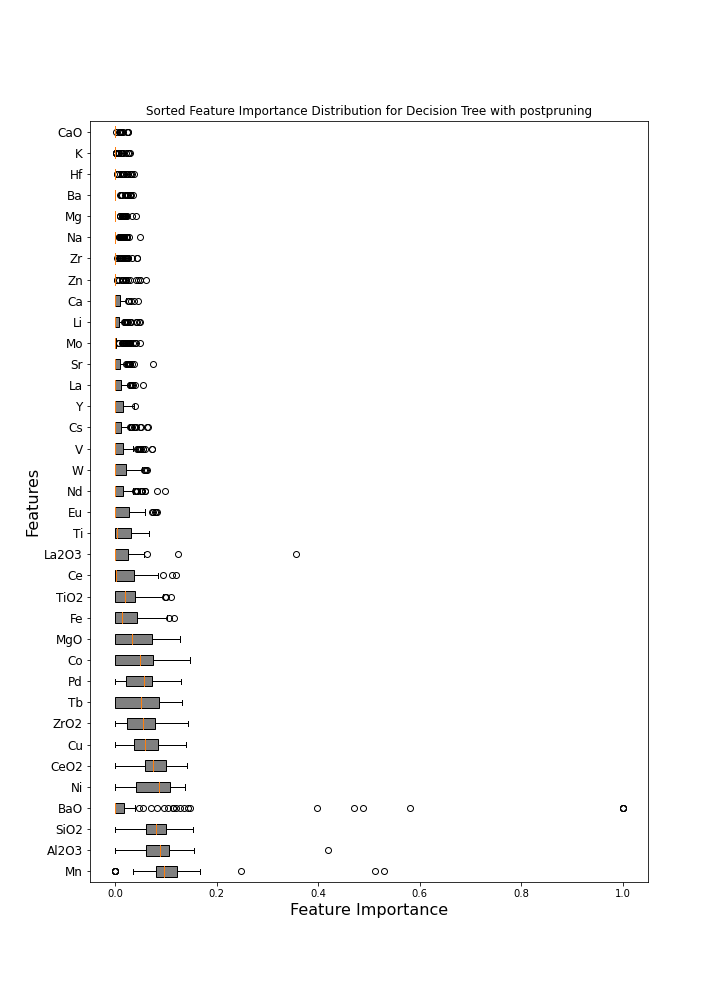}
        \caption{}
        \label{fig:right_dt_post}
    \end{subfigure}
    \caption{Feature importance analysis for decision tree with postpruning models}
    \label{fig:FI_4_DT_post}
\end{figure}

\FloatBarrier
\clearpage
\subsection{Random forest models}

\begin{table}[htbp!]
\centering
\caption{Hyperparameters Tuned in the random forest models}
\label{tab:hyperparameters_rf}
\begin{tabular}{@{}lll@{}}
\toprule
Hyperparameter            & Description                                                         & Range       \\ \midrule
\texttt{max\_depth}       & Maximum depth of each tree in the forest.                           & 1--10       \\
\texttt{n\_estimators}    & Number of trees in the forest.                                       & 50--500     \\
\texttt{min\_samples\_split} & Minimum number of samples required to split an internal node.     & 2--20       \\
\texttt{min\_samples\_leaf}  & Minimum number of samples required to be at a leaf node.          & 1--20       \\ \bottomrule
\end{tabular}
\end{table}

\begin{figure}[htbp!]
    \centering
    \begin{subfigure}{0.48\textwidth}
        \centering
        \includegraphics[width=\linewidth]{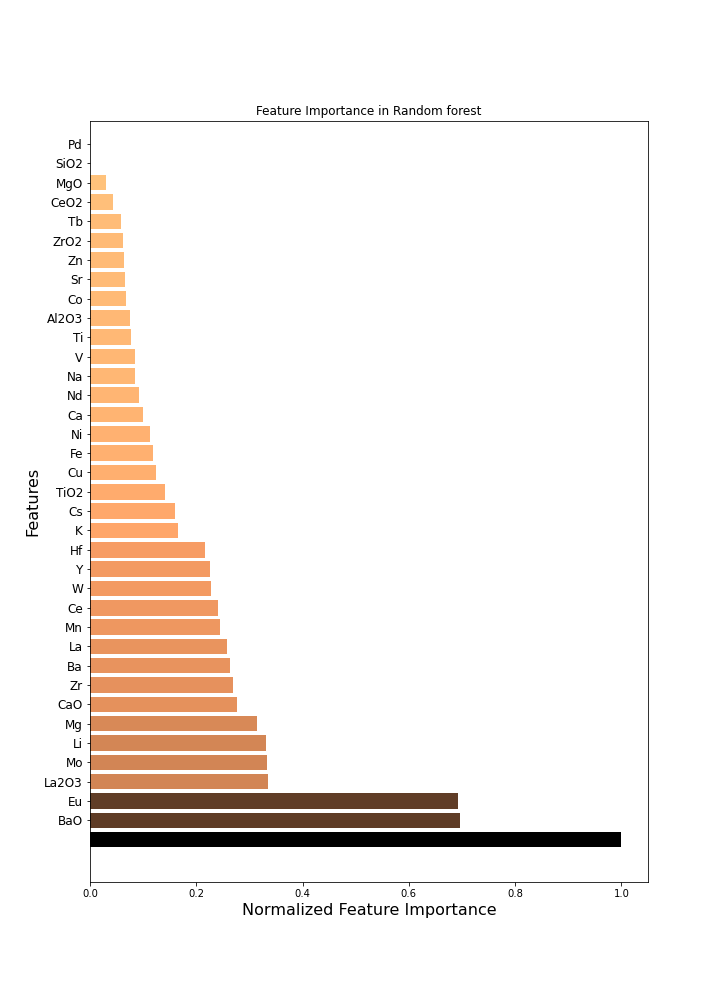}
        \caption{}
        \label{fig:left_rf}
    \end{subfigure}%
    \vspace{0.1cm}
    \begin{subfigure}{0.48\textwidth}
        \centering
        \includegraphics[width=\linewidth]{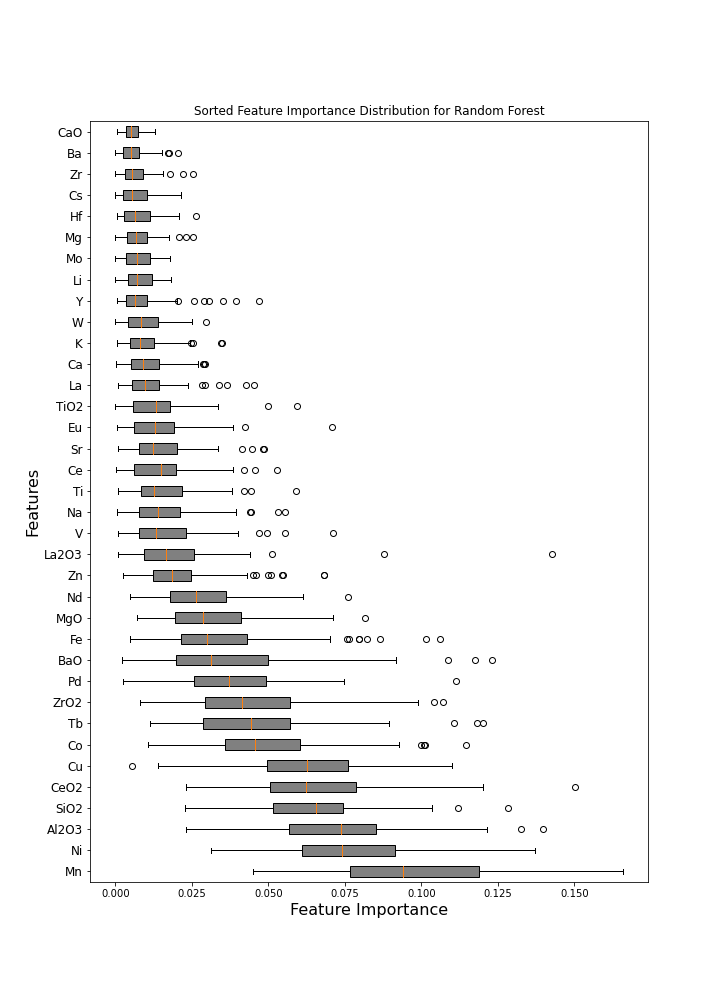}
        \caption{}
        \label{fig:right_rf}
    \end{subfigure}
    \caption{Feature importance analysis for random forest models}
    \label{fig:FI_4_rf}
\end{figure}

\FloatBarrier
\clearpage
\subsection{XGBoost models}\label{sec:xgb_theory}

\begin{table}[htbp!]
\centering
\caption{Hyperparameters Tuned in the XGBoost models}
\label{tab:hyperparameters_xgb}
\begin{tabular}{@{}lll@{}}
\toprule
Hyperparameter      & Description                                                        & Range                 \\ \midrule
\texttt{max\_depth}        & Maximum depth of each tree.                                        & 1--10                 \\
\texttt{learning\_rate}    & Step size shrinkage used to prevent overfitting.                  & $10^{-5}$ to $10^{0}$ \\
\texttt{n\_estimators}     & Number of trees in the ensemble.                                   & 50--500               \\
\texttt{reg\_alpha}        & L1 regularization term on weights (increases sparsity).           & $10^{-5}$ to $10^{0}$ \\
\texttt{reg\_lambda}       & L2 regularization term on weights (smoothens weights).            & $10^{-5}$ to $10^{0}$ \\ \bottomrule
\end{tabular}
\end{table}

\begin{figure}[htbp!]
    \centering
    \begin{subfigure}{0.48\textwidth}
        \centering
        \includegraphics[width=\linewidth]{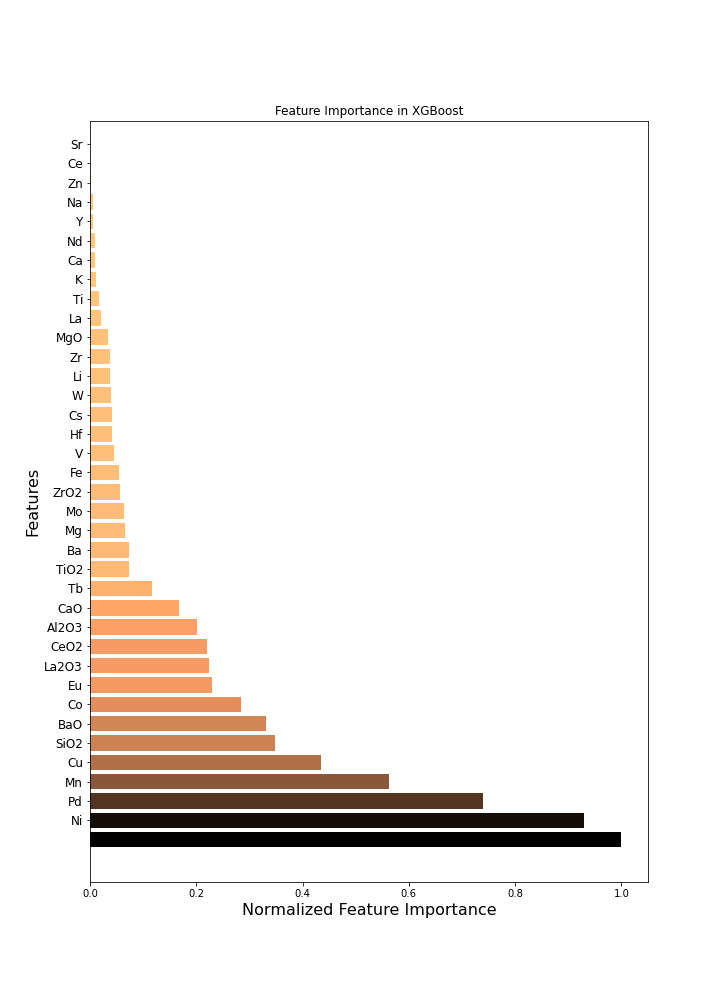}
        \caption{}
        \label{fig:left_xgb}
    \end{subfigure}%
    \vspace{0.1cm}
    \begin{subfigure}{0.48\textwidth}
        \centering
        \includegraphics[width=\linewidth]{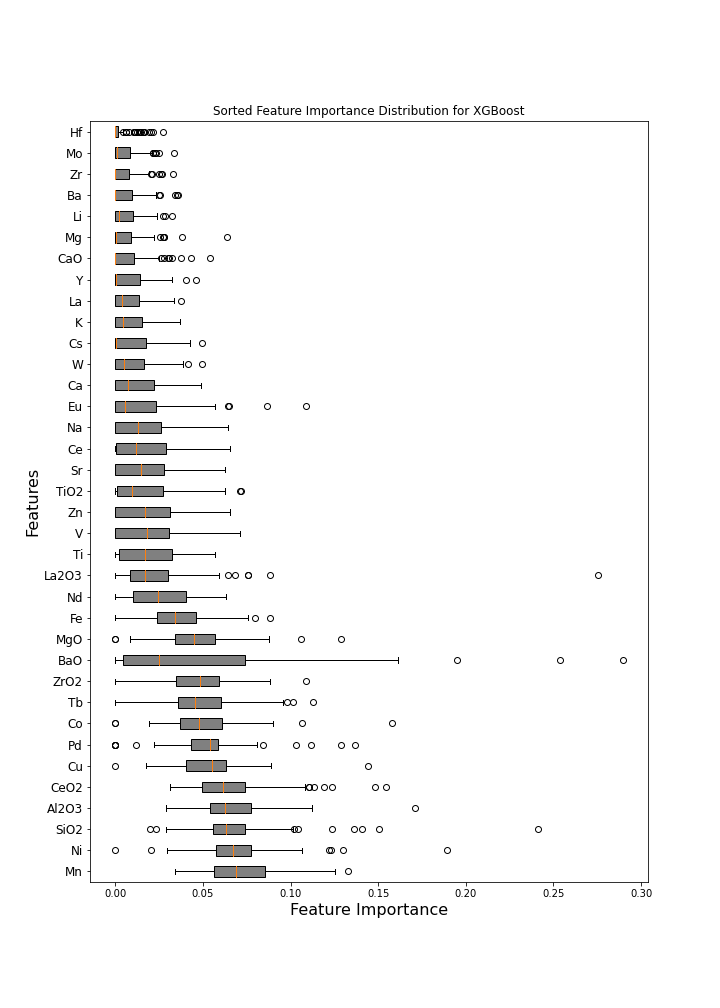}
        \caption{}
        \label{fig:right_xgb}
    \end{subfigure}
    \caption{Feature importance analysis for XGBoost models}
    \label{fig:FI_4_xgb}
\end{figure}

\FloatBarrier
\clearpage
\subsection{Logistic regression models}
\begin{figure}[htbp!]
    \centering
    \begin{subfigure}{0.48\textwidth}
        \centering
        \includegraphics[width=\linewidth]{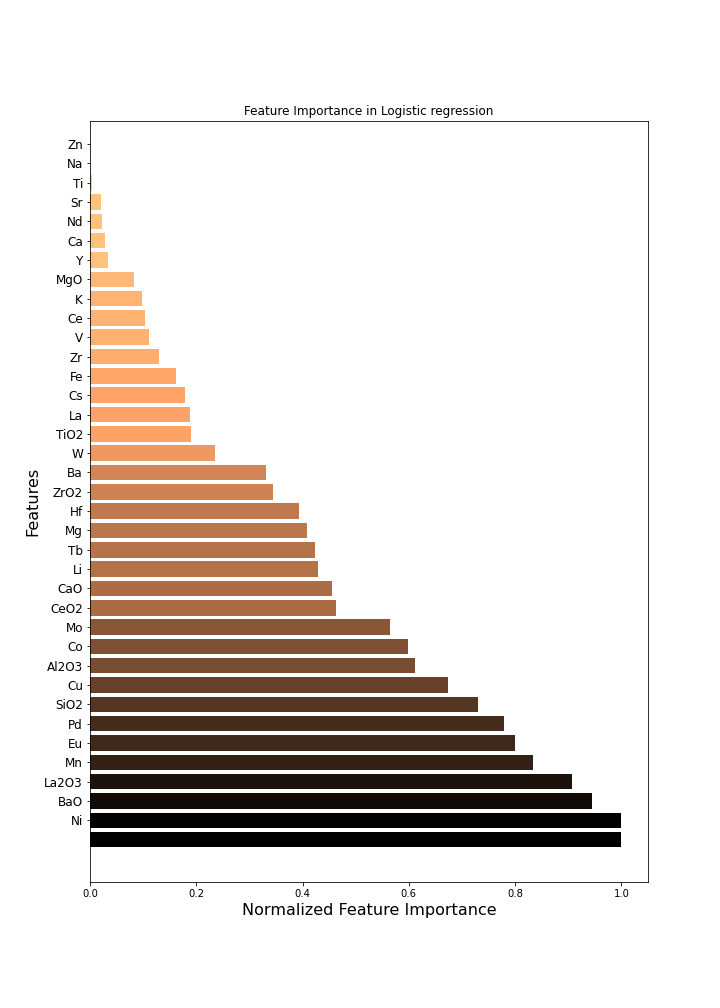}
        \caption{}
        \label{fig:left_lr}
    \end{subfigure}%
    \vspace{0.1cm}
    \begin{subfigure}{0.48\textwidth}
        \centering
        \includegraphics[width=\linewidth]{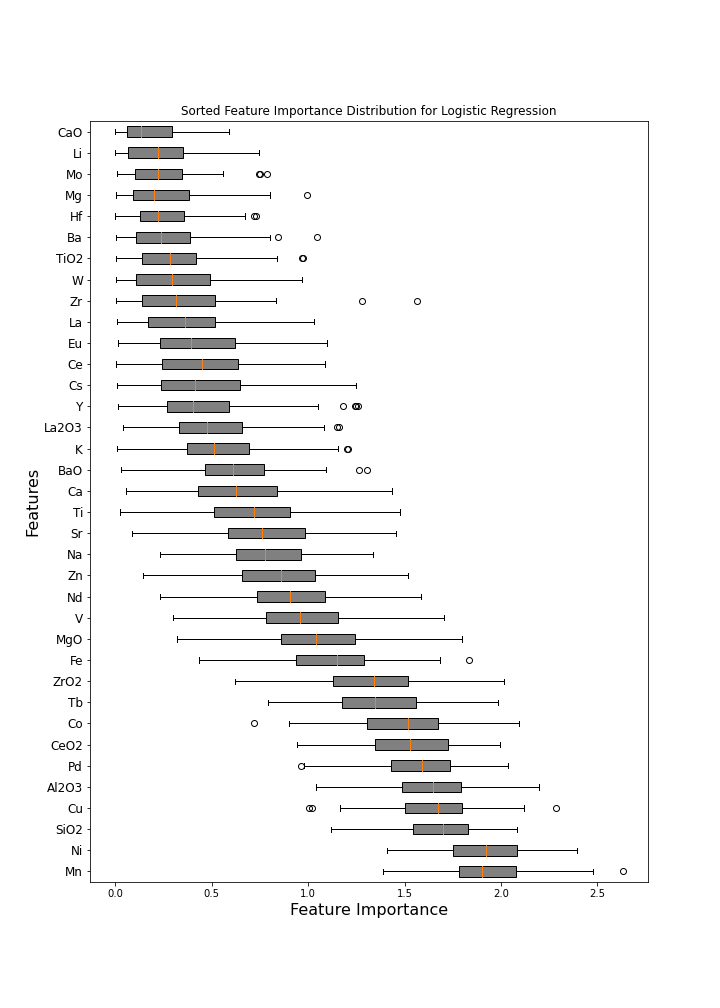}
        \caption{}
        \label{fig:right_lr}
    \end{subfigure}
    \caption{Feature importance analysis for logistic regression models}
    \label{fig:FI_4_lr}
\end{figure}

\FloatBarrier
\clearpage
\subsection{Neural networks models}

\begin{table}[htbp!]
\centering
\caption{Hyperparameters Tuned for the neural network models}
\label{tab:hyperparameters_nn}
\begin{tabular}{@{}lll@{}}
\toprule
Hyperparameter      & Description                                                        & Range                 \\ \midrule
\texttt{num\_layers}        & Number of hidden layers                                        & 2--4                 \\
\texttt{layer\_sizes}    & Number of neurons per layer  & \{16, 32, 36, 64, 128\} \\
\texttt{dropout\_rate}     & Probability for neuron dropout.                                   & \{0, 0.1\}               \\
\texttt{learning\_rate}    & Parameter update factor used to control training speed.                  & 0.001 \\
\texttt{weight\_decay}        & L2 weight decay used for optimizer      & 0.001 \\
\end{tabular}
\end{table}

\begin{figure}[htbp!]
    \centering
    \begin{subfigure}{0.48\textwidth}
        \centering
        \includegraphics[width=\linewidth]{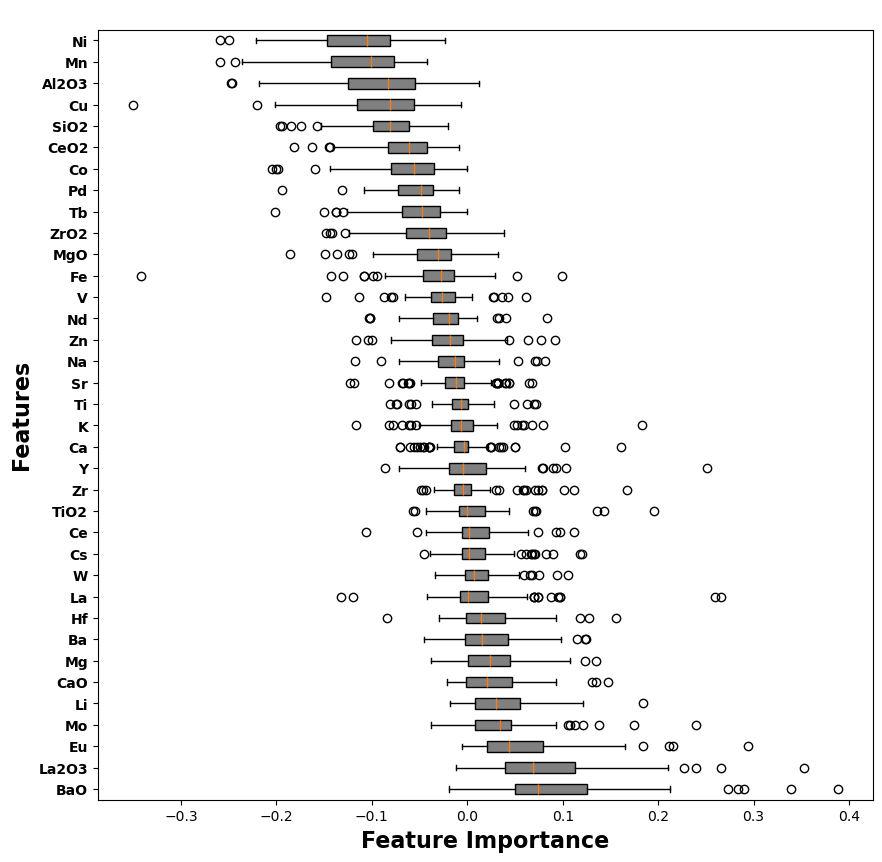}
        \caption{}
        \label{fig:left_nn_2}
    \end{subfigure}%
    \vspace{0.1cm}
    \begin{subfigure}{0.48\textwidth}
        \centering
        \includegraphics[width=\linewidth]{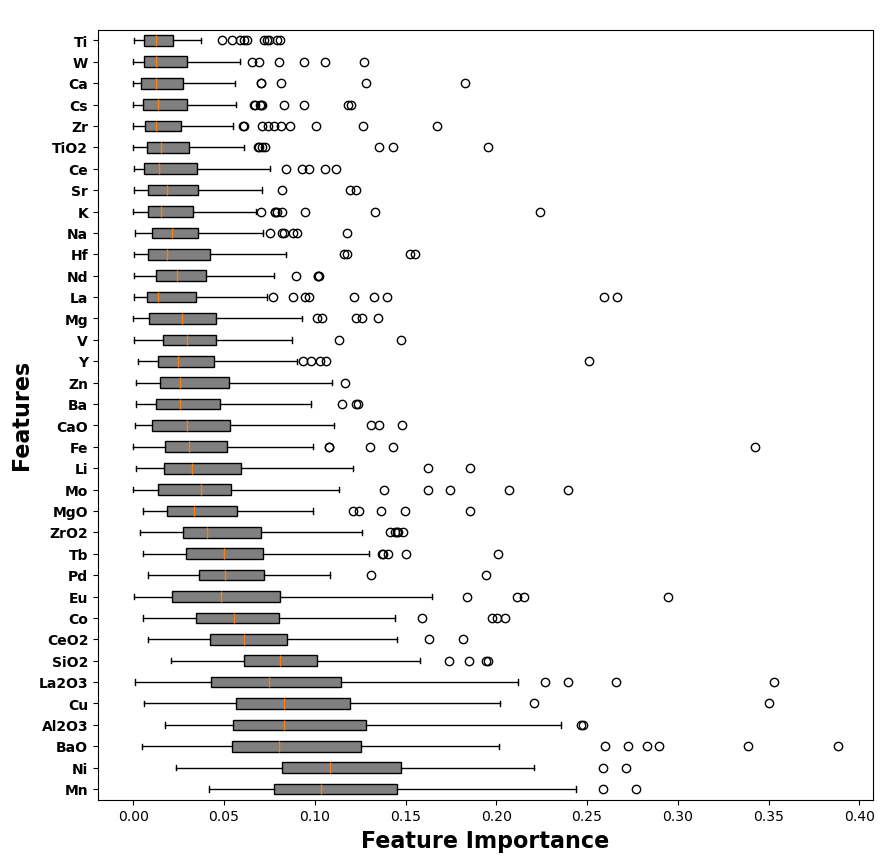}
        \caption{}
        \label{fig:right_nn_2}
    \end{subfigure}
    \caption{Feature importance analysis for neural networks models. (a) displays the distribution of feature importances absolute values. (b) illustrates the positive and negative contributions.}
    \label{fig:FI_4_nn_2}
\end{figure}
\FloatBarrier
\clearpage

\subsection{Feature importance of all models}
\vspace{-1em}
\begin{figure}[htbp!]
    \centering
    \includegraphics[width=0.5\textwidth]{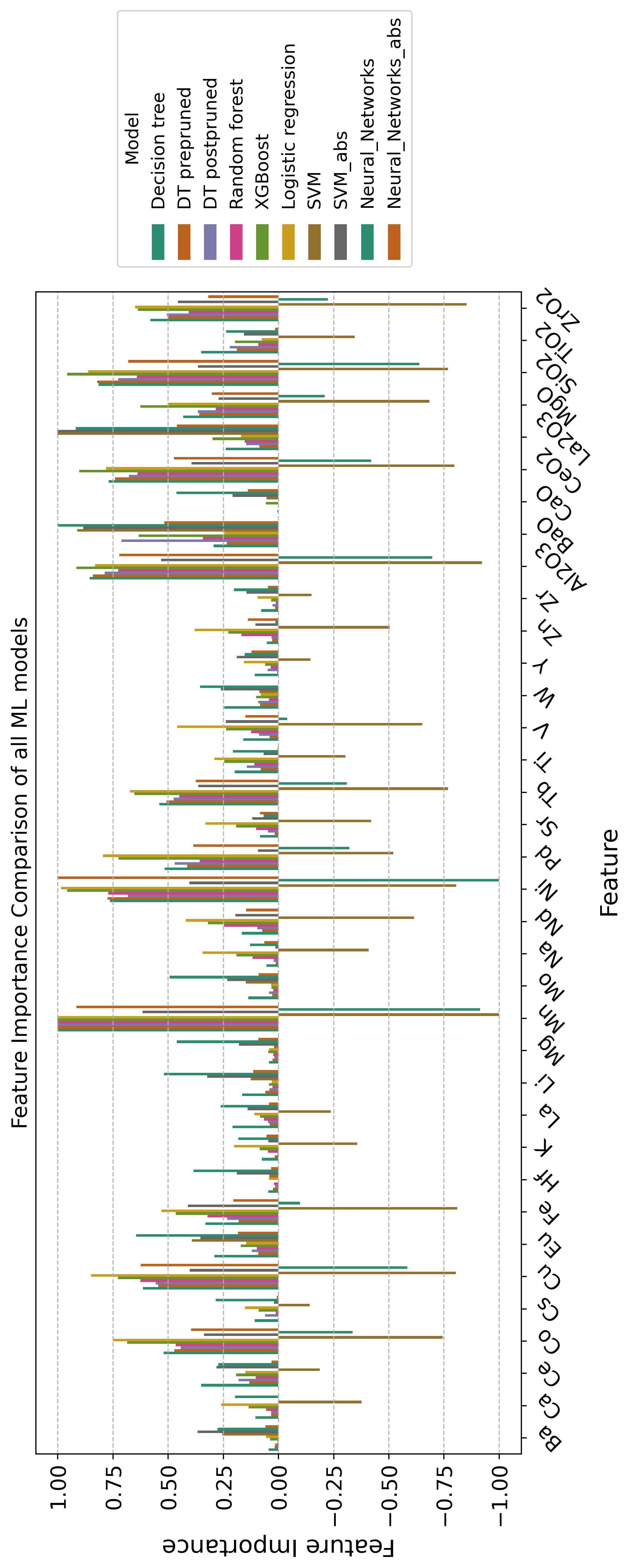}
    \caption{The normalized feature importances of all of the discussed machine learning models, averaged over 100 training and test splits of data}
    \label{fig:FI_all_models}
\end{figure}

\clearpage

\section{Single catalyst explanation with LRP}\label{sec:single_sample_lrp}
Besides providing class-aware explanations, the ability to produce explanations on a single sample basis is another property that separates LRP from the other explanation methods used in this work.

Having access to relevance scores for the components of each individual catalyst enables a chemist to examine in detail how much each component has contributed to the classification of a specific catalyst as high- or low-yield. 
Figure~\ref{fig:LRP single cat pos} visualizes the single catalyst feature relevances for a specific selection of samples, obtained using LRP from a neural network model. In order to automatically select interesting samples that are outliers in some way, we separated the samples into true positives, true negatives, false positives, and false negatives. From each of these four groups, we selected the samples that had the highest and lowest classifier scores for their predicted class. 
\begin{figure}[htbp]
    \centering
    \includegraphics[width=0.8\textwidth]{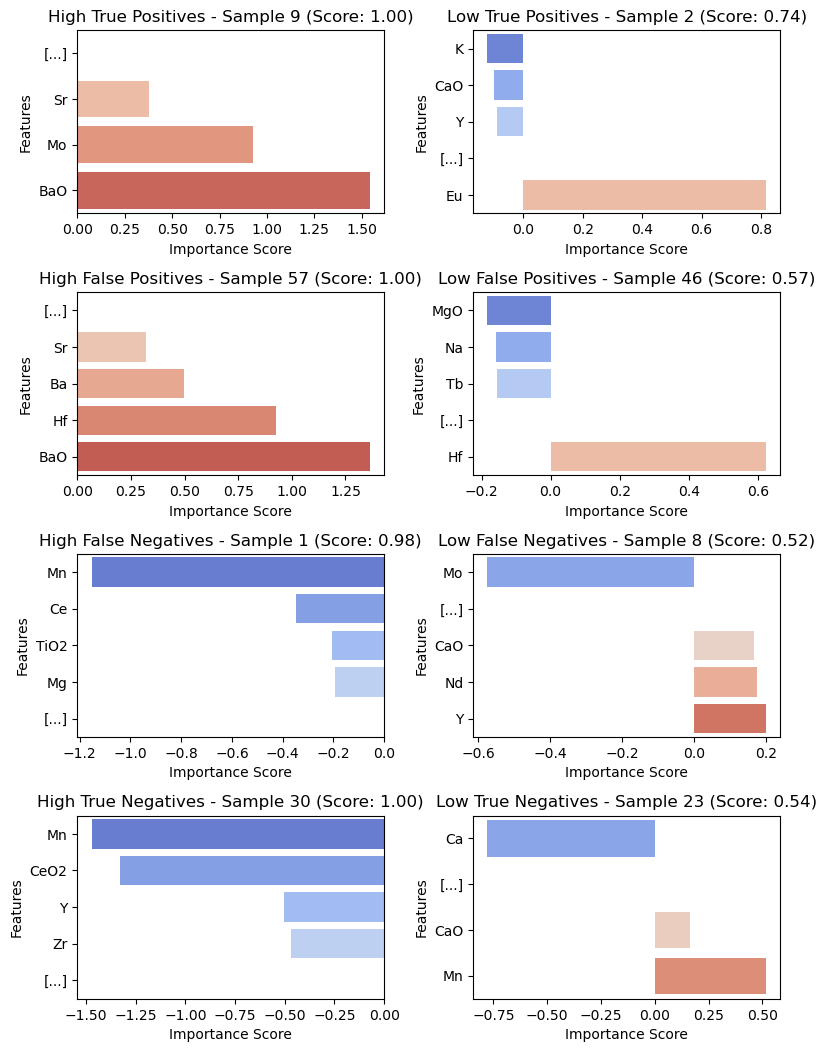}
    \caption{LRP for individual samples of catalysts, in all four categories of correct and incorrect predictions (TP, FP, FN, TN). From each category, two samples are demonstrated, one with a high confidence score and the other one with a low confidence score, for this prediction. }
    \label{fig:LRP single cat pos}
\end{figure}

We see that the relevances are distributed proportionally to the classifier scores, with high-confidence high- or low-yield samples having predominantly positive or negative relevance, respectively, whereas the samples with low classifier confidence have a mixture of positive and negative relevance, where the overall relevance adds up to a value close to zero.

It needs to be noted that since OCM reactions are very sensitive to the specific process condition, and this information is not present in the dataset, a more detailed interpretation of the single sample LRP explanations of this reaction cannot be made easily.

\clearpage

\section{Fisher-Z transformation}\label{sec:Fisher_Z}

Given the nature of Pearson correlation coefficients, which range between -1 and 1, direct arithmetic operations such as averaging can be misleading due to the non-linearity of the scale. To address this issue, we applied the Fisher Z transformation~\cite{fisher1915frequency} to the correlation coefficients. This transformation converts the correlation coefficients to a scale where they can be averaged and compared more meaningfully. The Fisher Z transformation is defined as:

\begin{equation}
    Z = \frac{1}{2} \ln\left(\frac{1 + r}{1 - r}\right)
\end{equation}

where $ r $ is the Pearson correlation coefficient. Special cases where $ r = 1 $ or $ r = -1 $ result in $ Z $ being set to positive or negative infinity, respectively, to maintain mathematical correctness.

Following the transformation, we calculated the average of the Fisher Z-transformed values for each unique pair of models. This allows us to quantify the average level of agreement between each pair of models, considering all feature importances. Subsequently, we converted these average Fisher Z values back to the original correlation scale using the inverse Fisher Z transformation, thereby obtaining mean correlation coefficients that are comparable across different model pairs.

\FloatBarrier
\clearpage

\section{Effect of periodic table group features}\label{sup:group_feats}

The dataset introduced in ~\cite{nguyen2021learning} included periodic table group features in addition to the elements and supports in order to improve the performance of their model. However, in many cases, the group information and element features are highly correlated (See Figure~\ref{fig:dataset_corr}), since most of the groups have only one element represented in the data (groups 3, 5, 7, 8, 9, 11 and 12 represented respectively by the following elements \ch{Y}, \ch{V}, \ch{Mn}, \ch{Fe}, \ch{Co}, \ch{Cu}, and \ch{Zn}).
As shown in Figure~\ref{fig:new Accuracy and F1 distribution}, whether we include the periodic system group information as separate features or not, the average performance of the model obtained using our evaluation framework is not affected. This is valid for all of the models in this work. However, the inclusion of these highly correlated features causes superfluous information, which makes the explanations highly difficult to disentangle.

\begin{figure}[htbp!]
    \centering
    \includegraphics[width=1\textwidth]{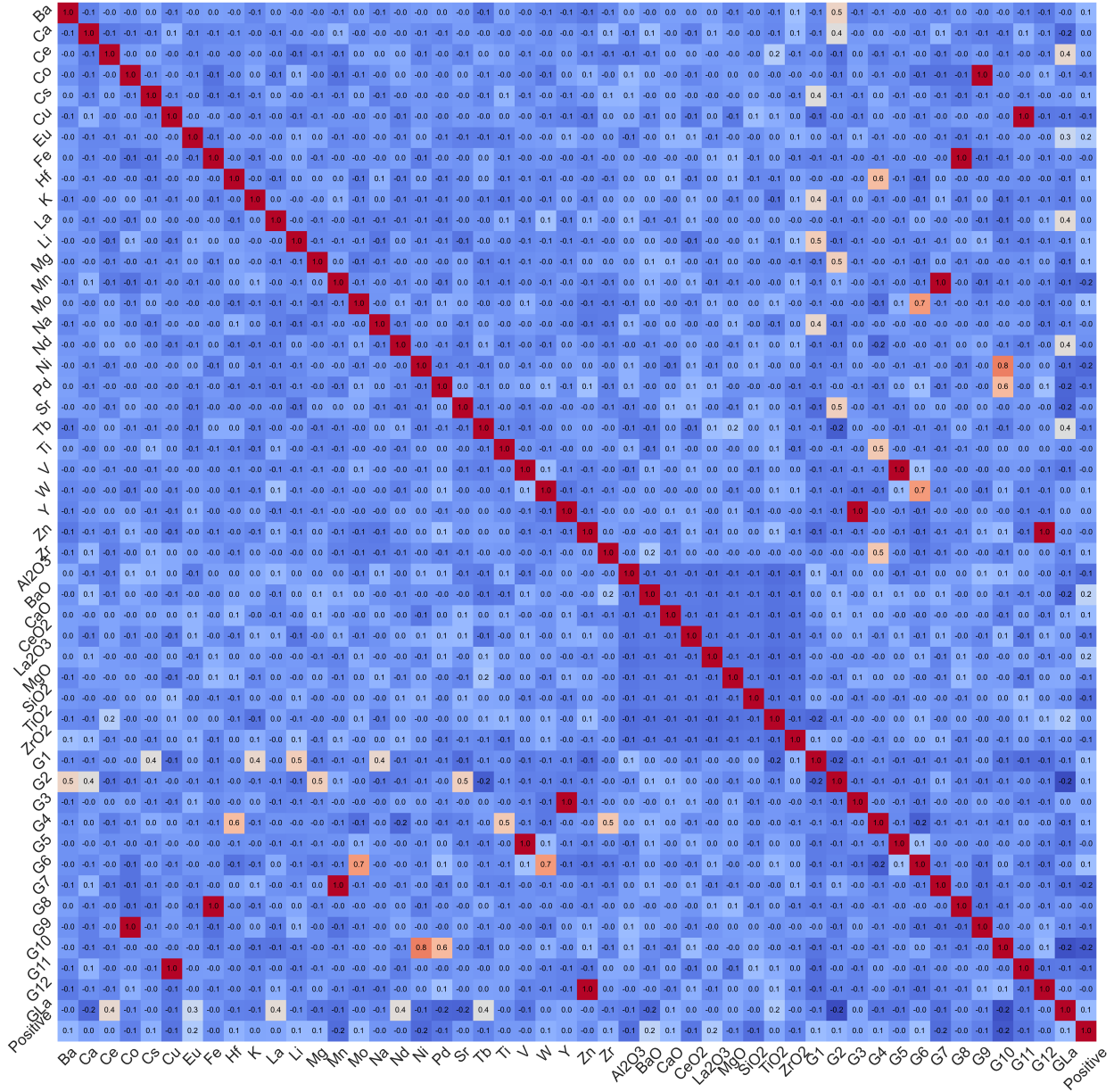}
    \caption{Correlation matrix between the features of OCM dataset by~\citeauthor{nguyen2021learning}}
    \label{fig:dataset_corr}
\end{figure}

As illustrated in Figure~\ref{fig:FI_LRP_w_group}, the relevance score of an element and its respective group does not always align with each other. For example, in samples 3 and 5, Group Lanthanide (\ch{GLa}) and \ch{La} have opposing signs. In Sample 32, \ch{Ce} and \ch{GLa} are both selected as the most important features with opposite signs. In sample 58 $Zr$ and $G4$, are having opposite signs while they are fully correlated.
\begin{figure}[htbp!]
    \centering
    \includegraphics[width=0.9\textwidth]{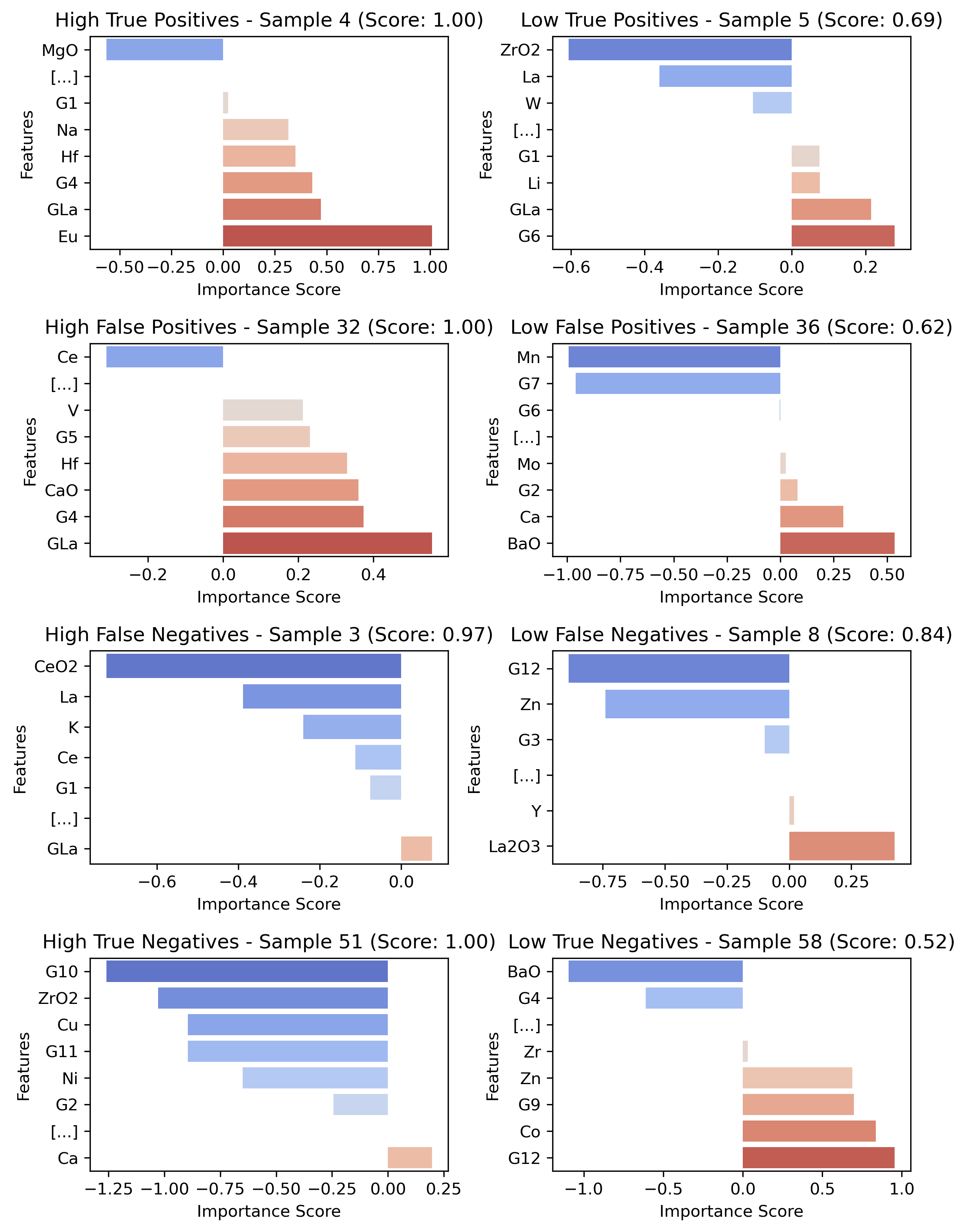}
    \caption{LRP analysis on single catalyst example with group information}
    \label{fig:FI_LRP_w_group}
\end{figure}

\FloatBarrier
This effect can also be seen when aggregating the feature importances for tree-based models. For example, $G10$ is represented by two elements in the dataset, namely Nickel (\ch{Ni}) and Palladium (\ch{Pd}). In the case of tree-based models, \ch{G10} consistently emerges as the most influential feature, while its individual elements, Nickel and Palladium, rank as the least important features.

\begin{figure}[htbp!]
    \centering
    \includegraphics[width=0.9\textwidth]{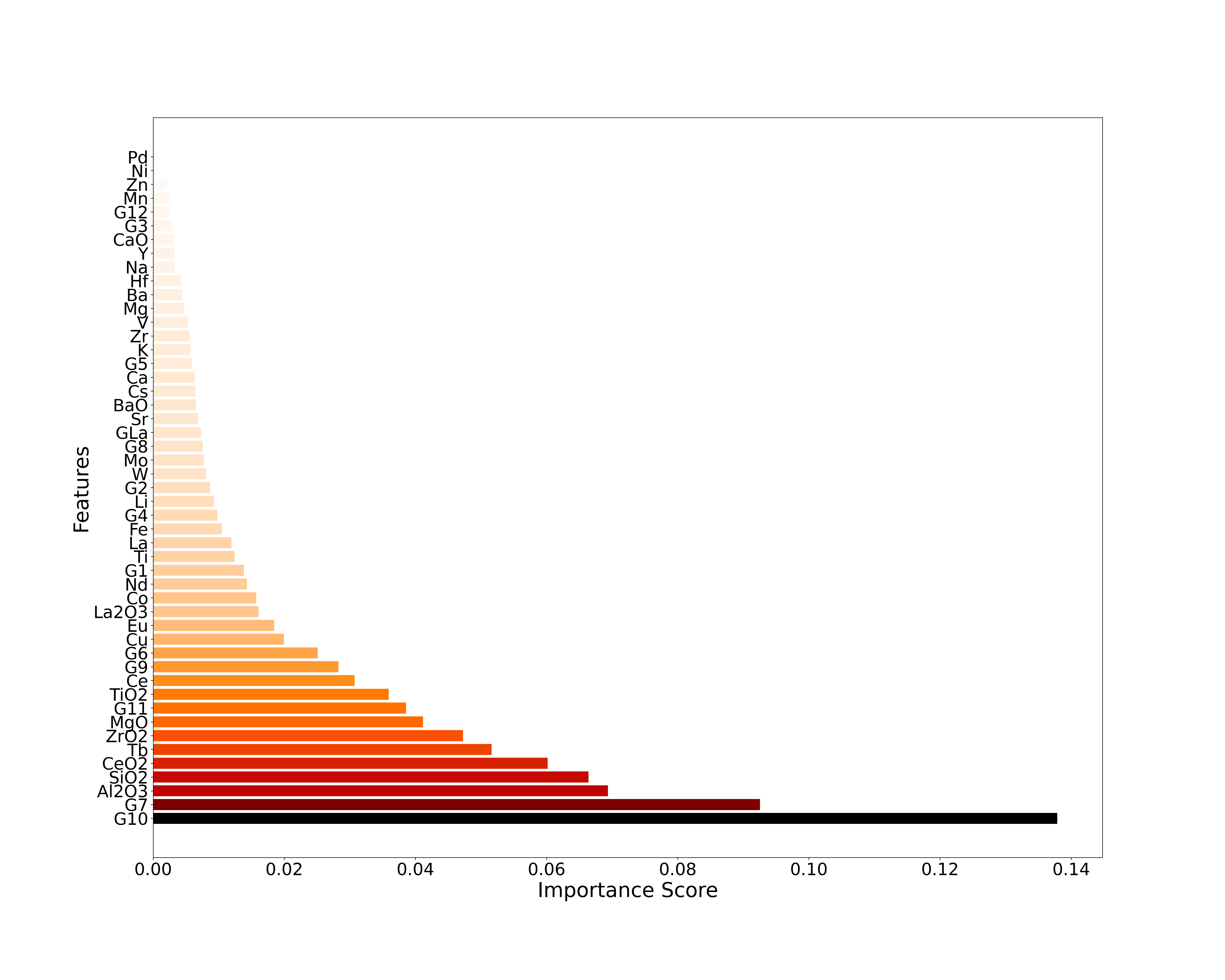}
    \caption{Average feature important analysis between tree-based models, with periodic system group information}
    \label{fig:FI_treemodels_w_group}
\end{figure}

\clearpage

\end{suppinfo}
\clearpage
\FloatBarrier
\bibliography{achemso-demo}

\end{document}